\documentclass[article]{JHEP}
\usepackage{epsfig}
\relax
\renewcommand{\theequation}{\arabic{section}.\arabic{equation}}
\def\be{\begin{equation}}
\def\ee{\end{equation}}
\def\bs{\begin{subequations}}
\def\es{\end{subequations}}
\newcommand{\een}{\end{subequations}}
\newcommand{\ben}{\begin{subequations}}
\newcommand{\beq}{\begin{eqalignno}}
\newcommand{\eeq}{\end{eqalignno}}

\def\bu{$\bullet \;\, $} 
\def\s{\sigma}

\def\l{\lambda}
\def\m{\mu}
\def\n{\nu}
\def\gn{G_{N}}
\def\a{\alpha}
\def\b{\beta}

\def\d{\partial}
\def\p{\partial}

\def\sp{\;\;\;,\;\;\;}
\def\r{\rho}

\def\e{\epsilon}

\newcommand\fverb{\setbox\pippobox=\hbox\bgroup\verb}
\newcommand\fverbdo{\egroup\medskip\noindent%
                        \fbox{\unhbox\pippobox}\ }
\newcommand\fverbit{\egroup\item[\fbox{\unhbox\pippobox}]}
\newbox\pippobox

\def\beq{\begin{equation}}
\def\eeq{\end{equation}}
\def\d{\delta}

\def\4R{{{}^{(4)}R}}

\def\K5{{\kappa}}
\def\K52{{\kappa^2}}

\newcommand{\ii}{i}
\newcommand{\jj}{j}

\newcommand{\da}{\dot{a}}
\newcommand{\db}{\dot{b}}
\newcommand{\dn}{\dot{n}}
\newcommand{\dda}{\ddot{a}}
\newcommand{\ddb}{\ddot{b}}

\newcommand{\pa}{a^{\prime}}
\newcommand{\pb}{b^{\prime}}
\newcommand{\pn}{n^{\prime}}
\newcommand{\ppa}{a^{\prime \prime}}

\newcommand{\ppn}{n^{\prime \prime}}
\newcommand{\fda}{\frac{\da}{a}}
\newcommand{\fdb}{\frac{\db}{b}}
\newcommand{\fdn}{\frac{\dn}{n}}
\newcommand{\fdda}{\frac{\dda}{a}}
\newcommand{\fddb}{\frac{\ddb}{b}}

\newcommand{\fpa}{\frac{\pa}{a}}
\newcommand{\fpb}{\frac{\pb}{b}}
\newcommand{\fpn}{\frac{\pn}{n}}
\newcommand{\fppa}{\frac{\ppa}{a}}

\newcommand{\fppn}{\frac{\ppn}{n}}

\def\square{\large\hbox{{$\sqcup$}\llap{$\sqcap$}}}
\def\hre#1#2{\href{http://arxiv.org/abs/#1/#2}{[ArXiv:#1/#2]}}
\def\hspi#1#2{\href{http://www.slac.stanford.edu/spires/find/hep/www?irn=#1}{#2}}

\title{Holography and Brane-bulk Energy Exchange}
\author{
Elias Kiritsis\\
CPHT, Ecole Polytechnique,
 91128, Palaiseau, FRANCE\\
  UMR du CNRS 7644.\\
  ~\\
and\\
~\\
Department of Physics, University of Crete\\
71003 Heraklion, GREECE\\
~\\
{\tt webpage: www.cpht.polytechnique.fr/cpth/kiritsis/} }

\preprint{\hepth{0504219} \\ CPHT RR 062.11.04}

\abstract{The five-dimensional description of generalized Randall-Sundrum cosmology is  mapped via holography to a generalization of the Starobinsky model.
This provides a holographic dual description of the cosmological brane-bulk energy exchange processes studied previously. Some simple solutions are presented in four dimensions.
}

\begin{document}

\tableofcontents

\newpage
\section{Introduction}

Branes and brane-worlds have provided novel contexts for studying both particle physics and cosmology.
\footnote{See \cite{report} for a recent review. Aspects of brane-world cosmology or particle physics are also reviewed in \cite{oreport}-\cite{oreportf}.} 

Recent observational data, demand an explanation for the presence and magnitude of dark energy.
They confirm the presence of an early inflationary era. This  demands natural initial conditions and an incorporation of the model in the mainstream of particle physics.
They also confirm the presence of dark matter, by leave its nature up for speculation.

Brane-world cosmology, motivated by string theory,  is one of several attempts to explain the observational data. Several new ideas have been proposed in this context. They include Randall Sundrum localization
\cite{rs} and its associated cosmology \cite{binetruy}, brane-world inflation \cite{dt}, mirage cosmology \cite{mirage}, variable speed of light realizations \cite{vsl}, brane/anti-brane inflation \cite{bb}
brane induced gravity \cite{big} and rolling tachyon cosmology \cite{tachyon}.

Brane-world models introduce a distinction between fields living on the brane and fields 
propagating in the bulk. Some of the SM fields are in the former class. The graviton is always in the latter class.

Since the early studies of brane-world models, it was appreciated that important constraints, as well as 
new effects are linked to interactions between brane and bulk fields, \cite{AADD}. 

Such interactions in the cosmological setting have been first investigated in
 \cite{he1}-\cite{he5}.    
 A systematic study of brane-bulk energy exchange was initiated in \cite{kkttz}.
 The full equations, projected on the brane were analyzed, and an approximation  was motivated. 
It was shown that acceleration was generic in the case of inflow and the associated fixed points
mostly stable. Further analysis confirmed these expectations \cite{tetr} and showed also 
new solutions indicating tracking behavior between brane and bulk energy density \cite{rep2}. 
Several more works \cite{allpapers1}-\cite{allpapers5} analyzed new cases, or provided some exact solutions \cite{RSr1,RSr2}.

An open question that mars attempts to discuss brane-cosmological equations without the knowledge of the 
full bulk solution , is the admissibility (or regularity) of solutions.
As AdS/CFT has indicated this is related in the holographic dual to the non-trivial dynamics 
of the boundary theory.

The purpose of this paper is to formulate the holographic dual setup of the general brane-bulk energy exchange equations derived in the context of the RS cosmology in \cite{kkttz}.
Previous work in this direction analyzing special cases includes \cite{rsh}.
   
The structure of this paper is as follows.
In section \ref{s2} we review the derivation of the general cosmological equations in RS cosmology.
In section \ref{s3} we present the map between the simplest RS cosmology and that of the hidden strongly coupled conformally invariant gauge theory. We also clarify several related confusions that exist in
literature. 
In section \ref{s4} we generalize the discussion to arbitrary cosmologies and relate them to arbitrary 
hidden, non-conformal gauge theories interacting withy observable matter. 
Some simple examples are described in section \ref{s5}.
In section \ref{s6} we speculate on bringing the holographic description of gravity to its logical conclusion
by advocating a gauge-theory description of four-dimensional gravity.

Appendix \ref{apa} contains our conventions and some useful geometrical formulae. 
In appendix \ref{anomaly} we review conformal anomalies in four-dimensions that play a crucial role
in the gravity/gauge theory correspondence. In appendix \ref{char}  we investigate the effects of adding a Gauss-Bonnet term in the five-dimensional bulk theory in order to test the modifications to the holographically dual theory. Finally in appendix \ref{D} we investigate the effect of the extra $R^2$ term in the anomaly to the four-dimensional graviton loops, to the associated cosmology.

\section{\label{s2}The brane-world viewpoint}
\setcounter{equation}{0}

In this section we will review the general setup of generalized RS
cosmology following \cite{binetruy} and \cite{kkttz} for the most general case
that includes a non-trivial interaction with bulk.

We shall be interested in the model described by the action\footnote{We could in principle add also a
four-dimensional Einstein term localized on the brane as in \cite{big}. This has been analyzed in \cite{ktt1,ktt2}. We will neglect it here.}
\be
S=\int d^{5}x~ \sqrt{-g} \left( M^3 R -\Lambda_5 +{\cal L}_B^{mat}\right)
+\int d^{4} x\sqrt{-\hat g} \,\left( -V+{\cal L}_b^{mat} \right),
\label{001}
\ee
where $R$ is the curvature scalar of the 5-dimensional metric
$g_{AB}, A,B=0,1,...,5$,
$\Lambda_5$ is the bulk cosmological constant, and
${\hat g}_{\alpha \beta}$, with $\alpha,\beta=0,1,...,3$,
is the induced metric on the 3-brane.
We identify
$(x,z)$ with $(x,-z)$, where $z\equiv x_{5}$. However, following the conventions
of \cite{rs} we extend the bulk integration over the entire interval
$(-\infty,\infty)$.
The quantity $V$ includes the brane tension as well as
possible quantum contributions to the
four-dimensional cosmological constant.

We consider an ansatz for the metric of the form
\begin{equation}
ds^{2}=-n^{2}(t,z) dt^{2}+a^{2}(t,z)\zeta_{ij}dx^{i}dx^{j}
+b^{2}(t,z)dz^{2},
\label{metric}
\end{equation}
where $\zeta_{ij}$ is a maximally symmetric 3-dimensional metric.
We use $ k$ to parameterize the spatial curvature.

The non-zero components of the five-dimensional Einstein tensor are
\begin{eqnarray}
{G}_{00} &=& 3\left\{ \fda \left( \fda+ \fdb \right) - \frac{n^2}{b^2}
\left(\fppa + \fpa \left( \fpa - \fpb \right) \right) + k \frac{n^2}{a^2} \right\},
\label{ein00} \\
 {G}_{\ii\jj} &=&
\frac{a^2}{b^2} \zeta_{ij}\left\{\fpa
\left(\fpa+2\fpn\right)-\fpb\left(\fpn+2\fpa\right)
+2\fppa+\fppn\right\}
\nonumber \\
& &+\frac{a^2}{n^2} \zeta_{ij} \left\{ \fda \left(-\fda+2\fdn\right)-2\fdda
+ \fdb \left(-2\fda + \fdn \right) - \fddb \right\} - k \zeta_{ij},
\label{einij} \\
{G}_{05} &=&  3\left(\fpn \fda + \fpa \fdb - \frac{\dot{a}^{\prime}}{a}
 \right),
\label{ein05} \\
{G}_{55} &=& 3\left\{ \fpa \left(\fpa+\fpn \right) - \frac{b^2}{n^2}
\left(\fda \left(\fda-\fdn \right) + \fdda\right) - k \frac{b^2}{a^2}\right\}.
\label{ein55}
\end{eqnarray}
Primes indicate derivatives with respect to
$z$, while dots derivatives with respect to $t$.

The higher-dimensional Einstein equations take the usual form
\beq
G_{AC}
= \frac{1}{2 M^3} T_{AC} \;,
\label{einstein}
\eeq
where $T_{AC}$ denotes the total energy-momentum tensor.

Assuming a perfect fluid on the brane and, possibly an additional energy-momentum
$T^A_C|_{m,B}$ in the bulk, we write
\begin{eqnarray}
T^A_{~C}&=&
\left. T^A_{~C}\right|_{{\rm v},b}
+\left. T^A_{~C}\right|_{m,b}
+\left. T^A_{~C}\right|_{{\rm v},B}
+\left. T^A_{~C}\right|_{m,B}
\label{tmn1} \\
\left. T^A_{~C}\right|_{{\rm v},b}&=&
\frac{\delta(z)}{b}{\rm diag}(-V,-V,-V,-V,0)
\label{tmn2} \\
\left. T^A_{~C}\right|_{{\rm v},B}&=&
{\rm diag}(-\Lambda_5,-\Lambda_5,-\Lambda_5,-\Lambda_5,-\Lambda_5)
\label{tmn3} \\
\left. T^A_{~C}\right|_{{\rm m},b}&=&
\frac{\delta(z)}{b}{\rm diag}(-\rho, p, p, p,0),
\label{tmn4}
\end{eqnarray}
where $\rho$ and $ p$ are the energy density and pressure on the brane, respectively.
The
behavior of $T^A_C|_{m,B}$ is in general complicated in the presence
of flows, but we do not have to specify it further at this point.

We wish to solve the Einstein equations at the location
of the brane. We indicate by the subscript o the value of
various quantities evaluated on the brane.

Integrating equations (\ref{ein00}), (\ref{einij})
with respect to $z$ around $z=0$ gives the known
jump conditions
\begin{eqnarray}
a_{o^+}'=-a_{o^-}'  &=& -\frac{1}{12M^3} b_o a_o \left( V + \rho \right)
\label{ap0} \\
n'_{o^+}=-n_{o^-}' &=&  \frac{1}{12M^3} b_o n_o \left(- V +2\rho +3 p\right).
\label{np0}
\end{eqnarray}

The other two Einstein equations (\ref{ein05}), (\ref{ein55})
give
\begin{equation}
\frac{n'_o}{n_o}\frac{\dot a_o}{a_o}
+\frac{a'_o}{a_o}\frac{\dot b_o}{b_o}
-\frac{\dot a'_o}{a_o} =
\frac{1}{6M^3}T_{05}
\label{la1}
\end{equation}
\begin{equation}
\frac{a'_o}{a_o}\left(\frac{a'_o}{a_o}+\frac{n'_o}{n_o}\right)
-\frac{b^2_o}{n^2_o}\left( \frac{\dot a_o}{a_o} \left( \frac{\dot
a_o}{a_o}-\frac{\dot n_o}{n_o}\right) +2\frac{\ddot a_o}{a_o}\right) -
k\frac{b^2_o}{a^2_o} =-\frac{1}{6M^3}\Lambda_5 b^2_o + \frac{1}{6M^3}T_{55},
\label{la2}
\end{equation}
where $T_{0,5}, T_{5,5}$ are the $05$ and $55$ components of $T_{AC}|_{m,B}$
evaluated at the position of  the brane.
Substituting (\ref{ap0}), (\ref{np0})
in equations (\ref{la1}), (\ref{la2}) we obtain
\begin{equation}
\dot {\rho} + 3 \frac{\dot a_o}{a_o} (\rho +p)
= -\frac{2n^2_o}{b_o}
T^0_{~5}
\label{la3}
\end{equation}
\begin{eqnarray}
\frac{1}{n^2_o} \Biggl(
\frac{\ddot a_o}{a_o}
+\left( \frac{\dot a_o}{a_o} \right)^2
-\frac{\dot a_o}{a_o}\frac{\dot n_o}{n_o}\Biggr)
+\frac{k}{a^2_o}
=\frac{1}{6M^3} \Bigl(\Lambda_5 + \frac{1}{12M^3} V^2
\Bigr)
\nonumber \\
-\frac{1}{144 M^{6}} \left( V (3 p-\rho ) +
\rho (3 p +\rho) \right) - \frac{1}{6M^3}T^{5}_{~5}.
\label{la4}
\end{eqnarray}

Choosing a gauge with $b_o=1$, a time coordinate on the brane 
so that $n_o=1$ and renaming  $a_o\to a$ we can
integrate the equation above at the expense of introducing the
mirage radiation density\footnote{This is also known as dark radiation.} $\chi$

\be {{{\dot a}^2}\over
{a^2}}={1\over 144M^6}\rho^2+{ V\over 72 M^6} (\rho+\chi) - {
k\over{a^2}}+ \lambda \label{a} \ee \be
\dot{\chi}+4\,{{\dot a}\over
a}\,\chi=\left({\rho\over V}+1\right)2
T^0_{~5}-24{M^3\over V}{{\dot a}\over a}{T^5}_5, \label{chi}
\ee 
\begin{equation}
\dot {\rho} + 3 \frac{\dot a}{a} (\rho +p)
= -2
T^0_{~5}
\label{la33}
\end{equation}
where

\be
 \lambda={1\over 12 M^3}\left(\Lambda_5+{V^2\over 12M^3}\right)\label{l}
 \ee
  is the effective
cosmological constant on the brane.
The RS case corresponds to $ T_{05}=T_{55}=0$ and $\lambda=0$:
\be {{{\dot a}^2}\over
{a^2}}={1\over 144M^6}\rho^2+{ V\over 72 M^6} (\rho+\chi) - {
k\over{a^2}}\label{ars} \ee \be
\dot{\chi}+4\,{{\dot a}\over
a}\,\chi=0 \sp 
\dot {\rho} + 3 \frac{\dot a}{a} (\rho +p)
= 0
\label{la3rs}
\end{equation}

The general equations (\ref{a}), (\ref{chi}) and (\ref{la33}) describe the cosmological
evolution of the brane-universe driven by two energy densities. The first,
$\rho$, the observable energy density on the brane. The other is the mirage density
$\chi$ that is determined by the bulk dynamics. In the simple RS case 
it behaves like free radiation.

In the general setup, these two energy densities interact. There is observable energy loss due
to a non-zero $T_{05}$. There is also mirage energy transmutation due to
both $T_{05}$ and $T_{55}$ in (\ref{chi}).
In a sense, $T_{05}$ controls the energy transfer between $\rho$ and $\chi$, namely brane and bulk.
$T_{55}$ provides a ``self-interaction" term to the bulk energy evolution.

Some issues of brane-bulk interaction were discussed in 
Depending on the nature of the bulk theory there may be very interesting cosmological solutions
to the equations (\ref{a})-(\ref{la33}).
Non-trivial solutions to these equations incorporate a non-trivial interaction between 
the brane energy density $\rho$ and the bulk energy density $\chi$. 
They were analyzed in detail first in \cite{kkttz}.
There, an approximation was motivated for a large class of solutions, namely neglecting $T_{55}$ from the equations. Furthermore $T_{05}$ was phenomenologically parameterized as a function of the observable
energy density $\rho$. Point-like behavior is characteristic of the scaling region around RS type solutions.
The solutions found had many interesting aspects:

(a) It turns out that accelerating fixed points, which are stable, 
are a generic property of the cosmological
equations in the case of inflow.

(b) Tracking solutions where  the mirage energy tracks the 
observable energy density \cite{rep2}

Further analysis and ans\"atze have been made in \cite{he1}-\cite{he5}.
An exact solution was also found, describing the radiation of energy from the brane
in the RS case \cite{he5},\cite{RSr1},\cite{RSr2}. In such cases it could be shown 
that $T_{55}$ becomes relevant  only close to a big-bang or a big-crunch singularity \cite{report},\cite{rep2}.

Such cosmological solutions may be relevant for explaining present or past cosmological acceleration in the universe. It is therefore appropriate to put such solutions at a more solid footing. 
This is the reason that we will try to provide a dual view of this system, using recent ideas of AdS/CFT
correspondence \cite{mald,review} and its generalizations .

\section{\label{s3}RS cosmology from the AdS/CFT correspondence}

In this section we will provide a link between the RS setup via the ADS/CFT
correspondence to a cut-off strongly coupled conformal gauge theory, coupled to
four-dimensional gravity \cite{g0}-\cite{g3}. 
We parameterize $AdS_5$
with Poincar\'e coordinates\footnote{These are not global coordinates
 but they will suffice for our exposition.
Global coordinate systems can be found in \cite{review}.},
\be 
ds^2 = {{\ell}^2\over
r^2}(dr^2-dt^2+d\vec x^2) 
\label{poi}\ee

 The flat RS space-time (RS brane)
corresponds to the region $R_1$ ($r>1$) of $AdS_5$ , cutoff at $r=1$
(position of the brane). The ultraviolet part ($r<1$) is replaced with
region $R_2$, a $Z_2$ copy of $R_1$.
The RS brane can be put at any other point of AdS, $r=r_0$, 
provided it is not the boundary, $r_0\not=0$,
with strictly equivalent physical results. This is due to the conformal symmetry of $AdS_5$
that acts by scaling the Poincar\'e coordinate $r$. 
 
Therefore, we expect to relate the RS setup to the cutoff AdS/CFT correspondence. To
do this, we will review below the IR regularization of the gravitational
(string) theory on $AdS_5$.

\subsection{Cut-off AdS$_5$/CFT$_4$}

Maldacena has conjectured \cite{mald} that string theory on $AdS_5\times
S^5$ with N units of four-form flux is dual to $N=4$ SU(N) Super
Yang-Mills theory. The gauge theory has two dimensionless parameters: the
number of colors N and the 't Hooft coupling $\lambda=g^2_{YM}N$. They are
related to the five-dimensional Planck scale M, the $AdS_5$ length $\ell$
and the string scale $l_{s}$ as

\be (4\pi)^2~ M^3\ell^3=2N^2 \sp \ell^2 =\sqrt{\lambda}~ l_s^2\ee

Quantum effects are suppressed on the string theory side when $N\to
\infty$. The geometry is macroscopic and the stringy effects suppressed
when $\lambda \to \infty$.

 The gravitational (string) theory on
$AdS_5\times S^5$ has as observables the path integral as a function of
the sources $\phi_i$ (at the $AdS_5$ boundary $r=0$) of the bulk fields
$\Phi_i$ \footnote{We will, until further notice, neglect the $S^5$ part
of the space. Our discussion is more general, and applies also 
to other AdS$_5$ related conformal theories.}

\be \phi_i(x_i)=\lim_{r\to 0} \Phi_i(r,x^i) \ee

\be Z_{\rm string}[\phi_i]=\int D\Phi_i~ e^{-S_{bulk}} \ee

There is a one-to-one correspondence between the bulk fields $\Phi_i$ and
single-trace operators $O_{i}$ of the boundary gauge theory. AdS/CFT
correspondence boils down to \cite{ren1,gkp}

\be Z_{\rm string}[\phi_i(x)]=e^{-W_{CFT}(\phi_i)}\equiv \langle ~
e^{~\sum_i\int d^4 x ~\phi_i(x)~ O_i(x)~}~\rangle \label{ads}\ee
where the average on the second hand side is in the boundary gauge theory.

As it stands, both sides of (\ref{ads}) are ill-defined without
regularization. The gravitational side has IR divergences due to the
infinite volume of AdS$_5$, visible as the fields approach the boundary
$r=0$. The gauge theory side has UV divergences, because the composite
gauge-invariant operators, $O_i$ are divergent, and the short-distance
divergences must be subtracted. There is a one-to-one correspondence
 between the IR divergences of the bulk theory and the UV divergences 
 of the boundary theory.
Both sides must be cutoff, and then
renormalized in the same way.

The renormalization on the gravitational side was described in
\cite{ren1}-\cite{ren9}. It involves cutting-off the theory before the boundary (at
$r=\e$), renormalizing the sources on that surface, and subtracting the
singular contributions from $S_{bulk}$. A convenient general
parametrization of the metric, due to Fefferman and Graham \cite{FG}, is as
 \footnote{The radial coordinate here is equal to the square of the Poincar\'e one in (\ref{poi}).}
\be 
{G_{MN}dx^Mdx^N\over
\ell^2}={dr^2\over 4r^2} +{g_{ij}(x,r)\over r}dx^idx^j 
\label{fg1}
\ee 
where 
\be
g_{ij}(x,r)=g_{(0)ij}(x) +rg_{(2)ij}(x)+r^2g_{(4)ij}(x)+r^2\log r ~
h_{(4)ij}(x)+{\cal O}(r^6)
\label{fg2}\ee

In the classical gravity limit (where, $N\to \infty$ and $\lambda\to
\infty$, where $\lambda$ is the 't Hooft coupling of the gauge theory),
this amounts to 

\be S_{bulk}\to S_{AdS}=S_{EH_5}+S_{GH_4}- S_{\rm counter}
\ee 
where
\be S_{EH_5}=M^3\int_{r\geq \e} d^5x \sqrt{-G}~ \left[R_{(5)}+{12\over
\ell^2}\right] \ee is the bulk (five-dimensional) Einstein-Hilbert action
and bulk cosmological constant, while $S_{GH_4}$ is the standard
Gibbons-Hawking boundary term \cite{GB}

\be S_{GH_4}=2M^3\int_{r=\e} d^4 x \sqrt{-\gamma}~ K \ee with $\gamma$ the
induced metric on the boundary and $K$ the trace of the second fundamental
form. The action $S_{\rm counter}$ localized at the boundary provides the
counter-terms required for the total action $S_{AdS}$ to be finite when
$\e\to 0$. It is  given by\footnote{These authors use Euclidean signature, 
and curvature conventions so that the AdS curvature is positive.
We have translated their results into our conventions.}  \cite{ren7},\cite{ren9}

\be S_{\rm counter}=\int d^4 x\sqrt{- g_{(0)}} \left[{6\over \e^2}-{1\over
2}\log \e\left((Tr g_{(2)})^2-Tr (g_{(2)})^2\right)\right] \ee

We may now use the following relations to trade $g_{(0)}$ with $\gamma$
the induced metric at the cut-off boundary $r=\e$

\be 
{\sqrt{- g_{(0)}}\over \e^2\sqrt{-\gamma}}=1-{\e\over
2}Tr[g_{(0)}^{-1}g_{(2)}]+{\e^2\over
2}[(Tr(g_{(0)}^{-1}g_{(2)}))^2+Tr(g_{(0)}^{-1}g_{(2)})^2]+{\cal
O}(\e^3)
\ee
\be 
Tr g_{(2)}={1\over 6\e}\left[-R[\gamma]+{1\over
2}\left(R_{ij}[\gamma]R^{ij}[\gamma]-{1\over 6}R[\gamma]^2\right)+{\cal
O}(R[\gamma]^3)\right] 
\ee
\be 
Tr(g_{(2)}^2)={1\over 4\e^2}\left[R_{ij}[\gamma]R^{ij}[\gamma]-{2\over
9}R[\gamma]^2 +{\cal O}(R[\gamma]^3)\right] 
\ee 
The terms cubic in the
curvatures above give vanishing contributions.

Using the relations above we can rewrite the counter-terms in terms of the
induced metric at the cut-off surface as

\be 
S_{\rm counter}=S_{0}+S_{1}+S_{2} \label{counter}
\ee 
\be 
S_0=6{M^3\over \ell}\int_{r=\e} d^4 x\sqrt{- \gamma} \sp S_1=-{M^3\ell\over
2}\int_{r=\e} d^4 x\sqrt{- \gamma}~ R[\gamma] \label{s01}
\ee
\be  
S_2={\log \e\over 4}{M^3\ell^3}\int_{r=\e} d^4 x\sqrt{- \gamma}~
\left[R_{ij}[\gamma]R^{ij}[\gamma]-{1\over
3}R[\gamma]^2\label{s2}\right]
\ee

The last term is cutoff-dependent and is responsible for the conformal
anomaly\footnote{We discuss conformal anomalies in four dimensions in appendix 
\ref{anomaly}.} \cite{ren3}. It is  in a frame with the scheme-dependent anomaly 
coefficient being $b=0$. In the general
b-frame, where the anomaly is given by (\ref{conf}),  we must add to
the counter-term action (\ref{counter})
also the term

\be S_b=-{b\over 6} \int d^4x \sqrt{-\gamma} ~R[\gamma]^2 \label{sb}\ee

Thus, in a general scheme, the counter-term action is

\be S_{\rm counter}=S_{0}+S_{1}+S_{2}+S_{b} \label{counter1}\ee

We may now do a rescaling of the radial variable (conformal transformation) to map $r=\e$ to $r=1$.
The only term that will change is the cutoff-depended term that is responsible for the conformal anomaly, $S_2$.

The precise form of AdS/CFT correspondence relation (\ref{ads}) is

\be Z_{\rm string}[\phi_i(x)]=\int_{r>1} D\Phi_i~
e^{-S_{EH_5}-S_{GH_4}+S_0+S_1+S_2+S_b}= e^{-W_{CFT}(\phi_i)}
\label{adsr}\ee

We may now generalize the arguments in \cite{hawk1},\cite{skenderis} in order to derive the
dual formulation of the RS setup. We add to the gravitational sector a matter sector localized on the $r=1$ boundary
(RS brane). The relevant action in the RS setup can
be written as

\be S_{RS}=S_{EH_5}+S_{GH_4}-2S_0+S_{\rm matter}\label{rs}\ee

Here $2S_0$ is the tension of the RS brane, and $S_{\rm matter}$ is the
action of the matter localized on the RS brane.

We may now write the RS path-integral as

\be Z_{\rm RS}[\phi_i,\chi_i]=\int_{R_1\cup R_2} ~ D\Phi_iD\chi_i~
e^{-S_{\rm RS}} \ee where as usual $\phi_i$ are the values of the bulk
fields $\Phi_i$ on the RS brane, and $\chi_i$ are the extra matter fields
on the brane. $R_1$ is the region $r>1$, while $R_2$ is the region $r<1$,
related by the $Z_2$ symmetry to $R_1$. Since the integrals over $R_i$ are
independent (and equal) we can replace them with an integral only on one
side, say $R_1$ 

 \be Z_{\rm RS}[\phi_i,\chi_i]=\int_{R_1} ~ D\Phi_iD\chi_i~ e^{-2S_
{EH_5}-2S_{GH_4}+2S_0-S_{matter} }\ee and using (\ref{adsr}) we finally
obtain

\be Z_{\rm RS}[\phi_i,\chi_i]=\int_{R_1} ~ D\Phi_i~
e^{-2W_{CFT}-2S_1-2S_2-2S_b }\int D\chi_i~ e^{-S_{\rm
matter}}\label{ads-rs}\ee

Therefore, the relevant action of the RS dual is

\be S_{\tilde{RS}} =S_{CFT}+S_{R}+S_{R^2}+S_{\rm matter}\ee with \be
S_{CFT}=2W_{CFT} \sp S_{P}=2S_1=-M_P^2~ \int ~ d^4 x\sqrt{- \gamma}~
R[\gamma] \label{action}\ee

\be S_{R^2}=2S_2+2S_b=b'\int d^4 x\sqrt{- \gamma}~
\left[R_{ij}[\gamma]R^{ij}[\gamma]-{1\over 3}R[\gamma]^2\right]-{b\over
3}\int d^4 x\sqrt{- \gamma}~R[\gamma]^2\ee with $M_P^2=M^3\ell$, and $b'$
is a scheme-dependent coefficient coming from the logarithmic counterterm, 
while $b$ is the (scheme-dependent)
b-coefficient of the conformal anomaly (see Appendix \ref{anomaly}).

The generalization  to the theory where there is no perfect fine-tuning 
of the bulk and brane vacuum energies, e.g. $\lambda$ in
(\ref{l}) involves adding a cosmological term in (\ref{action}).
The general final four-dimensional dual action is

\be S_{\tilde{RS}} =S_{CFT}+S_{\lambda}+S_{P}+S_{R^2}+S_{\rm matter}\ee with \be
S_{CFT}=2W_{CFT} \sp S_{P}=2S_1=-M_P^2~ \int ~ d^4 x\sqrt{- \gamma}~
R[\gamma] \label{actionl}\ee
with
\be
S_{\lambda}=-2\lambda\int d^4x \sqrt{-\gamma}
\ee

We may match now the parameters: In the four dimensional description we have
$M_P$ and $c=N^2/(8\pi)^2$. In the five dimensional description we have $M, V$ and
\be
 \ell=24M^3/V\sp \Lambda_5=-{V^2\over 12M^3}
 \ee
 We obtain
\be
M_P^2=24{M^6\over V}\sp c=\left(12{M^4\over V}\right)^3 = \left({M\ell\over 2}\right)^3
\label{iden}\ee
where $c$ the conformal anomaly is given in (\ref{n}).
Inversely,
 \be
 M^2={M_P^2\over 2 c^{1\over 3}}\sp V={3\over c}M_P^4=3\left({8\pi\over N}\right)^2 M_P^4
 \sp \ell={2\sqrt{2c}\over M_P}\sp \Lambda_5=-{3\over c\sqrt{2c}}M_P^5
\label{adscft} \ee
In particular, at large $N$, $M_P>> M$.

Our "derivation" of the gauge-theory dual description of the RS setup has
two subtleties.

The first is technical: The slicing of the space-time we have assumed to
derive the gauge-theory dual description is ok, for smooth space-times like
$AdS_5$, but maybe problematic for other space-times, in particular if
they contain classical singularities (like big-bang singularities for
example). If  such singularities are in the bulk we can imagine that the
slicing can be done without spoiling the argument.

The second is more essential: strictly speaking the RS five-dimensional
solution cannot be embedded simply into the IIB supergravity as required
for the advertised duality. We may start from the $AdS_5\times S^5$
solution, cut-off $AdS_5$ to generate the localized energy, but this
solution fails the self-duality conditions for the four-form.
In particular, they imply that there should also be some source in the
$S^5$ directions. Although it is possible that by distributing
symmetrically branes on $S^5$ may do the job, an exact solution is not
known.\footnote{We would like to thank K. Stelle for a discussion on this.}
In the five-dimensional language, such a solution would correspond to
turning on some of the other fields of the supergravity.

Our attitude here is to ignore this subtlety and take this approximate map as a motivation 
for taking the gauge theory description
seriously. In the discussion section we will present a framework where this seems motivated.

\subsection{Cosmological evolution in the holographic-dual description}

Our starting point are the equations of motion obtained by varying
(\ref{action}) with respect to the four-dimensional metric $\gamma$.

They are of the form
\footnote{The variations of the $R^2$ terms are presented in Appendix \ref{apa}.}
\be 
M_P^2 G_{\m\n}+\lambda \gamma_{\m\n}=T^{m}_{\m\n}+W_{\m\n}+Z_{\m\n} \label{4}
\ee
where
\be
T^{m}_{\m\n}={1\over \sqrt{-\gamma}}{\delta S_{\rm matter}\over \delta
\gamma^{\m\n}}\sp
W_{\m\n}={1\over \sqrt{-\gamma}}{\delta W_{CFT}\over \delta
\gamma^{\m\n}}\ee

\be Z_{\m\n}={1\over \sqrt{-\gamma}}{\delta S_{R^2}\over \delta
\gamma^{\m\n}}=b'\left[
-2\nabla^{\rho}\nabla_{\n}R_{\r\m}+{2\over
3}\nabla_{\m}\nabla_{\n}R+\square
R_{\m\n}+2{R_{\m}}^{\r}R_{\r\n} -{2\over
3}RR_{\m\n}+\right.\label{z}\ee
$$
\left.+{1\over 6}\gamma_{\m\n}(R^2-\square R-3R^{\r\s}R_{\r\s})\right]+{b\over
3}\left[2\nabla_{\m}\nabla_{\n}R-2RR_{\m\n}+{1\over
2}\gamma_{\m\n}(R^2-4\square R)\right]
$$
All stress tensors are conserved
\be
\nabla^{\m}T^m_{\m\n}=\nabla^{\m}W_{\m\n}=\nabla^{\m}Z_{\m\n}=0
\ee

Moreover, as explained in Appendix \ref{anomaly},  we have the following trace formulae\footnote{We neglect for the
moment the contributions to the trace due to the loops of the four
dimensional graviton. These will be treated in Appendix \ref{D}.}
\be
{W^{\m}}_{\m}=-2a~ C^2-2c~G+2b~  \square R \sp {Z^{\m}}_{\m}=-2b \square R
\ee

Defining the combined stress tensor
\be
V_{\m\n}=W_{\m\n}+Z_{\m\n}
\ee
we have
\be
 \nabla^{\m}V_{\m\n}=0\sp {V^{\m}}_{\m}=-2a~ C^2-2c~G
\ee

For a homogeneous cosmological background $ds^2=-dt^2+a^2 \zeta_{ij}dx^idx^j$
we parameterize the stress tensors as
\be T_{00}=\rho\sp T_{ij}=p~a^2\zeta_{ij} \ee
\be
V_{00}=\s \sp V_{ij}=\s_{p}~a^2\zeta_{ij} \ee

Then, the equations (\ref{4}) become

\be
M_P^2{\dot a^2\over
a^2}={1\over 3}(\rho+\s+\lambda)-{k\over a^2}\sp M_P^2\left({\dot a^2\over
a^2}+2{\ddot a \over a}\right)=-p-\s_p-{k\over a^2}+\lambda
\ee
\be
\Longrightarrow M_P^2{\ddot a \over a}=-{1\over
6}(\rho+3p+\s+3\s_p)+{\lambda\over 3}
\ee

The conservation equations
$\nabla^{\m}T_{\m\n}=\nabla^{\m}V_{\m\n}=0$ amount to
\be
\dot \rho+3{\dot a \over a}(\rho+p)=0\sp \dot \s+3{\dot a \over a}(\s+\s_p)=0
\ee

The conformal anomaly equation implies
\be
\s-3\s_p=48c{\ddot a \over a}\left[{\dot a^2\over a^2}+{k\over a^2}\right]\sp c={N^2\over (8\pi)^2}>>1
\label{n}\ee
where $N$ is the number of colors of the gauge theory.
Solving for $\s_p$ and substituting in the conservation equation
we obtain
\be
\dot \s +4H\s-48c~H(H^2+{k\over a^2})(\dot H+H^2)=0
\ee
with solution
\be
\s=\chi_{rad}+12c\left[H^2+{k\over a^2}\right]^2\sp \dot
\chi_{rad}+4H\chi_{rad}=0\Longrightarrow
\chi_{rad}={\chi_0\over a^4}\ee
It should be stressed that the positivity of the energy in the strongly coupled gauge theory sector
reads
\be
\s= \chi_{rad}+12c\left[H^2+{k\over a^2}\right]^2\geq 0
\label{pos}\ee

Therefore, the FRW equation is
\be
3M_P^2\left[H^2+{k\over a^2}\right]=12c\left[H^2+{k\over
a^2}\right]^2+\rho+\chi_{rad}+\lambda
\label{eeq}\ee
with solution
\be
H^2={M^2_P\over 8c}\left[1+\e\sqrt{1-{16c\over 3M_P^4}(\rho+\chi_{rad}+\lambda)}\label{sol}\right]-{k\over a^2}.
\ee
where $\epsilon=\pm 1$.
For $\e=1$ and $\rho+\chi_{rad}+\lambda=0$ this is essentially the Starobinsky solution \cite{star1}-\cite{star3}, 
used to generate inflation leading to 
\be H=H_0=\frac{M_P}{\sqrt{4c}}\sp V_s\equiv H_0^4={M_P^4\over 16c^2} \label{star}\ee
for a flat universe.
The inflation ends by the gradual growing of an extra higher-derivative term.
This is essentially the $b$-term in the anomaly. We have seen that a correct treatment of the
AdS/CFT map shows that such a term is absent for the RS-dual.
However, there will be higher terms in the curvature, that we have neglected here.
Such terms are suppressed at strong coupling by inverse powers of the 't Hooft coupling.
In appendix \ref{char}, we present the corrections, due to the Gauss-Bonnet term in the bulk.
Higher derivative terms are also expected in general.

We define the critical (maximal) energy density  
\be
E_0={3M_P^4\over 16c}
\label{critical}
\ee
and rewrite the FRW evolution as
\be
H^2+{k\over a^2}={H_0^2\over 2}\left[1+\e\sqrt{1-{\rho+\chi_{rad}+\lambda\over E_0}}\label{sol1}\right]
\ee

We will now rescale 
\be
\rho\to E_0\rho\sp \chi_{rad}\to E_0\chi_{rad}\sp \lambda\to E_0\lambda\sp t\to t/H_0\sp k\to k H_0^2
\label{scaling}
\ee
so that all variables become dimensionless.  
(\ref{eeq}) becomes
\be
4\left[H^2+{k\over a^2}\right]=4\left[H^2+{k\over
a^2}\right]^2+\rho+\chi_{rad}+\lambda
\label{eeqnew}\ee
with solution
\be
H^2+{k\over a^2}={1\over 2}\left[1+\e\sqrt{1-(\rho+\chi_{rad}+\lambda)}\right]\label{sol1a}
\ee
By further rescaling the scale factor $a$, $k$ can be set to $\pm1,0$ as usual.

For small densities, $\rho,\chi_{rad},\lambda<<1$,
\be
H^2={1+\e\over 2}-{\e\over 4}(\rho+\chi_{rad}+\lambda)-{k\over a^2}-
{\e\over 16}(\rho+\chi_{rad}+\lambda)^2+....
\label{lq}\ee
The smooth branch, $\e=-1$, matches the five dimensional evolution (\ref{ars})-(\ref{la3rs}).
The RS parameters can be then written in terms of the gauge theory parameters as
\be
V=2E_0\sp M^3={E_0\over 3H_0}\sp \Lambda_5=-E_0H_0
\label{lq1}\ee

The late time evolution, captured by the linear terms, matches well the gravitational description of section 2 in the smooth branch ($\e=-1$).

 However, in the regime 
where the quadratic terms in (\ref{lq}) become comparable to the linear terms,
the two evolutions diverge.
The four-dimensional gauge theory description has an upper critical density, namely $E_0$.
This bound is not visible in the five-dimensional gravitational description.

\subsection{A Simple cosmological solution of the conformal gauge theory}

We will consider $p=w\rho$ observable matter without curvature (k=0) and $\lambda=\chi=0$ 
\be
\rho=\left({a_0\over a}\right)^{3(1+w)}\sp H^2={1\over 2}
\left[1+\e\sqrt{1-{\rho}}\right]
\ee
where $a_0$ is the critical scale factor where $\rho=1$.

The cosmological solution is
\be
{2\over \sqrt{1+\e\sqrt{1-{\rho}}}}+{1\over \sqrt{2}}\log\left[{\sqrt{2}+\sqrt{1+\e\sqrt{1-{\rho}}}\over \sqrt{2}-\sqrt{1+\e\sqrt{1-{\rho}}}}\right]=
\ee
$$=
{3(1+w)}
{1\over \sqrt{2}}(t-t_0)+2+{1\over \sqrt{2}}\log{\sqrt{2}+1\over
\sqrt{2}-1}
$$
where we have parameterized the constant of integration so that at $t=t_0$, $\rho=1$.
$\rho$ decreases with time monotonically until it becomes zero at $t=\infty$.

In the Starobinsky branch, $\e=1$, the expansion is exponential at all times. 
In the smooth branch, $\e=-1$, the expansion is power-like.

\section{\label{s4}The general holographic case}

Once the conformal case is understood, the general non-conformal
 and interacting case is straight-forward.
 Modifying the bulk theory in the RS setup amounts to modifying the boundary gauge theory.
 Having an non-trivial $T_{05}$ amounts to including interactions between the 
 Strongly
Coupled Gauge Theory (SCGT) and the observable matter.
We will assume therefore, that the SCGT is perturbed away from
conformality and that there are interaction terms coupling it to
the matter theory.

The relevant action is
\be S_{general} =S_{\rm SCGT}+S_{R}+S_{R^2}+S_{\rm matter}+S_{\rm interaction}\ee
A potential vacuum energy is lumped into the SCGT action.

Integrating out the strongly coupled fields, replaces $S_{\rm SCGT}+S_{\rm
interaction}$ with $W_{\rm SCGT}$ which is a (non-local) functional of the
metric and the observable matter fields.

Finally the relevant action becomes

\be S_{general} =W_{\rm SCGT}+S_{R}+S_{R^2}+S_{\rm matter}\ee
The $R^2$ terms  are as before.

The Einstein equation now reads
\be
M_P^2 G_{\m\n}=T^m_{\m\n}+V_{\m\n}
\label{6}\ee
with
\be
T^{m}_{\m\n}={1\over \sqrt{-\gamma}}{\delta S_{\rm matter}\over \delta
\gamma^{\m\n}}\sp
V_{\m\n}={1\over \sqrt{-\gamma}}{\delta (W_{SCGT}+S_{R^2})\over \delta
\gamma^{\m\n}}\ee
where
\be
\nabla^{\m}T^m_{\m\n}=- \nabla^{\m}V_{\m\n}\equiv T\sp {V^{\m}}_{\m}=-2a~
C^2-2c~G+D ~R^2
\ee
with $D={\delta \over 90(4\pi)^2}$ and $\delta$ given in appendix \ref{anomaly}.
The extra contribution to the anomaly is possible if there are
non-conformally coupled scalars as well as from the graviton and gravitini.
We have $D<<c$. We will set here $D=0$. The case $D\not=0$ is analyzed in Appendix \ref{D}.

For a homogeneous cosmological background $ds^2=-dt^2+a^2 g_{ij}dx^idx^j$
we again parameterize the stress tensors as
\be T_{00}=\rho\sp T_{ij}=p~a^2g_{ij} \sp
V_{00}=\s \sp V_{ij}=\s_{p}~a^2g_{ij} \ee

Then, the Einstein equations (\ref{6}) become

\be
M_P^2{\dot a^2\over
a^2}={1\over 3}(\rho+\s)-{k\over a^2}\sp M_P^2\left({\dot a^2\over
a^2}+2{\ddot a \over a}\right)=-p-\s_p-{k\over a^2}
\ee
\be
\Longrightarrow M_P^2{\ddot a \over a}=-{1\over
6}(\rho+3p+\s+3\s_p)
\ee

The conservation equations
$\nabla^{\m}T_{\m\n}=-\nabla^{\m}V_{\m\n}=T$ amount to
\be
\dot \rho+3{\dot a \over a}(\rho+p)=-T\sp \dot \s+3{\dot a \over
a}(\s+\s_p)=T
\ee

The conformal anomaly equation implies
\be
\s-3\s_p=48c{\ddot a \over a}\left[{\dot a^2\over a^2}+{k\over
a^2}\right]+X
\ee
The term $X$ is collecting all classical and quantum breakings of the
conformal anomaly, due to masses and $\beta$-functions.
Before the SCGT integration $X$ has the general form
\be
X=\sum_{ij}(\beta^{(1)}_{ij}+R~\beta^{{(2)}}_{ij})O_i \Omega_j+(\beta^{(3)}_{ij}R_{\m\n}+\beta^{(4)}_{ij}g_{\m\n})O^{\mu}_i \Omega^{\nu}_j
\label{anomalyg}\ee
where $O_i$ are operators  in the SCGT and $\Omega_j$ are
operators in the matter theory.
We have also suppressed the space-time metric in (\ref{anomalyg}). 
Upon integration on the gauge SCGT fields
\be
X=\sum_i (B^{(1)}_i+B^{(2)}_i~R) ~\Omega_i+(B^{(3)}_i R_{00}-B^{(4)}_{i})~\Omega^0_{i} \sp B^{(I)}_j=\sum_j\beta^{(I)}_{ij}\langle 
O_i\rangle
\ee
The operators that survive above are quasi-scalar operators 
(singlets under the spacial rotation group).

Solving for $\s_p$ and substituting in the conservation equation
we obtain
\be
\dot \s +4H\s-48c~H\left(H^2+{k\over a^2}\right)(\dot H+H^2)=T+H~X
\ee
which upon defining the new density $\chi$ becomes
\be
\s=\chi+12c\left[H^2+{k\over a^2}\right]^2\sp \dot
\chi+4H\chi=T+H~X\ee

Consequently, the full system can be summarized into the FRW equation

\be
3M_P^2\left[H^2+{k\over a^2}\right]-12c\left[H^2+{k\over a^2}\right]^2=\rho+\chi
\label{frwg}
\ee
\be
\dot
\chi+4H\chi=T+H~X
\label{chia}\ee
\be
\dot \rho+3H(\rho+p)=-T
\label{rho}
\ee

Some comments are in order here.

\bu Standard dimensional arguments indicate that the four dimensional action will have 
up to $R^2$ terms but not more. The reason is visible in the possible counter-terms
 in the gravitational side.
 
\bu The proper treatment of the scheme-depended $\square R$ contribution of the anomaly indicates 
that it does not appear in the final four-dimensional cosmological equations.
Such terms have been used in the Starobinsky model to exit from inflation.
Their presence in this context can only be forced by fiat.

\subsection{Acceleration}

We may calculate the acceleration in this context:
\be
q=\dot H+H^2={\rho-3p+X-6M_P^2\left[H^2+{k\over a^2}\right]\over 6\left(M_P^2-8c\left[H^2+{k\over a^2}\right]\right)}={-(\rho+3p)+X-2\left(\chi+12c\left[H^2+{k\over a^2}\right]^2\right)\over 6\left(M_P^2-8c\left[H^2+{k\over a^2}\right]\right)}
\label{acc}\ee

Solving (\ref{frwg}) we obtain
\be
H^2={M^2_P\over 8c}\left[1+\e\sqrt{1-{16c\over 3M_P^4}(\rho+\chi)}\right]-{k\over a^2}.
\label{frwg1}\ee
so that 
\be
q=-\e{-(\rho+3p)+X-2\left(\chi+12c\left[H^2+{k\over a^2}\right]^2\right)\over 6M_P^2\sqrt{1-{16c\over 3M_P^4}(\rho+\chi)}}
\ee

In the smooth branch, $\e=-1$ due to the positivity condition (\ref{pos})
the only terms that can be positive, are the $\rho+3p$ combination for $w<-1/3$ and X.
We have acceleration iff
\be
 X>(\rho+3p)+2\left(\chi+12c\left[H^2+{k\over a^2}\right]^2\right)
 \ee
 
 In the Starobinsky branch $\e=1$ we obtain acceleration in exactly the opposite case 
 \be
 X<(\rho+3p)+2\left(\chi+12c\left[H^2+{k\over a^2}\right]^2\right)
 \ee
 Therefore, we always have an accelerating solution.

The best general intuition can be obtained by 
going back to the original energy and pressure variables for the hidden theory $\s$,$\s_p$
where we will separate only the conformal anomaly:
$\s_p\to \s_p$+anomaly.
Now $\s\geq 0$.
The equations become
\be
3M_P^2\left[H^2+{k\over a^2}\right]=\rho+\s\sp \dot \rho+3{\dot a \over a}(\rho+p)=-T
\label{new1}\ee

\be
\dot \s +3H(\s+\s_p)=T+48c~H\left(H^2+{k\over a^2}\right)(\dot H+H^2)
\label{new2}\ee
and we obtain for the acceleration
\be
q=\dot H+H^2=-{1\over 6M_P^2}{\rho+3p+\s+3\s_p\over 1-{8c\over 3M_P^4}(\rho+\s)}
\ee

 The denominator is due to the conformal anomaly.
It is positive in the smooth branch and negative in the Starobinsky branch.
In this form the conditions for acceleration are clear:

\bu In the smooth branch, $\rho+3p$ or $\s+3\s_p$
should be sufficiently negative.

\bu In the Starobinsky branch, any standard matter with $\rho+3p>0$ is accelerating.

In particular, the linearized fixed points, of \cite{kkttz} in the case of inflow, are forbidden here 
by the positivity of the energy $\sigma$. They are replaced by the Starobinsky branch, fixed points.

\subsection{Comparison with the five-dimensional gravitational equations}

This general system of equations above (\ref{frwg})-(\ref{rho}) should
 be compared with the ones derived earlier in the gravitational approach
\be H^2={1\over 144M^6}\rho^2+{ V\over 72 M^6} (\rho+\chi)+ \lambda \label{anew} \ee \be
\dot{\chi}+4H\chi=\left({\rho\over V}+1\right)2
T^0_{~5}-24{M^3\over V}H{T^5}_5\sp \dot\rho+3H(\rho+p)=-2{T^0}_5
 \label{chi1}
\ee

They can be matched in the leading terms by the identification
\be
T\simeq 2{T^0}_{5}\sp X\simeq-24{M^3\over V}{T^5}_{5}
\ee
The difference appears as before in the $\rho^2$ term and its $\rho
T_{05}$ avatar.
It should be noted that we may redefine the radiation density so that the equations match exactly
at the expense of redefining also the X density:
\be
\chi_{\rm 5d}=\chi_{4d}+12c \left(H^2+{k\over a^2}\right)^2-{c\over 6M_P^4}\rho^2
\ee
\be
{T^5}_{5}=-{V\over 24M^3}\left[X+ {\rho(\rho+3p)\over V}+4(12)^4{M^{12}\over V^3}  \left(H^2+{k\over a^2}\right)\left(\dot H+H^2\right)\right]
\ee
$$
=-{M_P\over 2\sqrt{2c}}\left[X+{c\over 3}{\rho(\rho+3p)\over M_P^4}+48c  \left(H^2+{k\over a^2}\right)\left(\dot H+H^2\right)\right]
$$
where we also need to use (\ref{iden}).

 \section{\label{s5}Simple explicit examples}
 
 In the previous section we have derived the general cosmological equations for the interaction of observable 
 matter with hidden theory.
 They are valid for any four dimensional hidden theory, which is assumed not observable (and therefore
 its degrees of freedom are integrated out.
 Although the motivation and starting point was the cosmology of five-dimensional brane worlds 
 and the ensuing brane-bulk energy exchange, one can take the four-dimensional point of view 
 as an independent definition of the cosmology. For some hidden theories that are strongly coupled and nearly conformal the cosmology resembles the five-dimensional RS cosmology.

As we will see here even after a  spontaneous breaking of conformal invariance and/or renormalization group flow the cosmological equations are very similar.
In this section, for purposes of illustration we will describe such examples. In these examples 
the coupling $T$ between the observable and the hidden sector remains null, but there can be a non-trivial contribution to the trace $X$ of the hidden stress-energy tensor. 
Realistic applications will be the subject of a future publication.

We will describe here cases which involve a single scalar perturbation of the hidden CFT.
In this case, the scalar is dual to an operator of dimension $\Delta$ of the CFT.
Its asymptotic expansion, parallel to the one in (\ref{fg1},\ref{fg2}) is
\be
\phi(r,x)\sim r^{{4-\Delta\over 2}}\left[\phi_0(x)+{\cal O}(r)\right] 
\ee
where the asymptotic value of the scalar , $\phi_0$ is the "coupling" constant.

The general expression for 
the nontrivial part of the trace $X$ is of the following form \cite{ren8}
\be
X=(\Delta-4)\phi_0~\langle O_{\Delta}\rangle+S(\phi_0)
\label{ex}\ee
 where $ \langle O_{\Delta}\rangle$ is its 
one-point function.

The Coulomb branch case \cite{cb,cb1} corresponds to giving an expectation value to the $\Delta=2$ chiral operator
$Tr[X^IX^J+X^JX^I]$ for I=1,J=2.
In this case 
\be
S(\phi_0)={1\over 2}\phi_0^2
\ee
in (\ref{ex}).
However, since there is no perturbation in the action, $\phi_0=0$ and finally $X=0$.
This is an example where conformal invariance is broken spontaneously because of the Higgs vev, but 
there is no non-trivial contribution to the trace.

The second example involves the GPPZ flow \cite{gppz}. This is a perturbation of the CFT with an operator
of dimension $\Delta=3$ which gives mass to the three chiral multiplets breaking supersymmetry $N=4\to N=1$.
Here 
\be
S(\phi_0)={1\over 2}\left[\phi_0\square \phi_0+{1\over 6}R\phi_0^2\right], \phi_0={\sqrt{3}\over \ell}
\ee
and it turns out that  $ \langle O_{\Delta=3}\rangle =0$.
Therefore $X={1\over 4\ell^2}R$ and the cosmological evolution equation
(\ref{chia}) becomes
\be
\dot\chi+4H\chi=-{3\over 2\ell^2}H\left(\dot H+2H^2+{k\over a^2}\right)
\ee
which can be solved as 
\be
\chi=\chi_{rad}-{3\over 4\ell^2}\left(H^2+{k\over a^2}\right)\sp
 \dot\chi_{\rm rad}+4H\chi_{\rm rad}=0
\ee
From (\ref{frwg}) this is equivalent to conserved dark radiation and a renormalized Planck scale

\be
\tilde M_P^2=M_P^2-{3\over 4\ell^2}=M_P^2\left(1-{6\pi^2\over N^2}\right)
\ee

\renewcommand{\theequation}{\arabic{section}.\arabic{subsection}.\arabic{equation}}
\section{\label{s6}Further thoughts on gravity, cosmology and gauge theory}
\setcounter{equation}{0}

In the previous sections, we derived a 4-dimensional description for the general brane-bulk 
interactions in 5 dimensions, at a cosmological setting.
The dual analogue replaces the physics of the 5d bulk with a strongly coupled
four-dimensional (hidden) gauge theory which may interact
 weakly with the observable matter sector
and four-dimensional gravity. The duality link is given by the bulk-boundary correspondence
in the form of the AdS/CFT cut-off correspondence.

 Its generalization will be to drop the strong coupling criterion for the gauge theory 
 and therefore consider the interaction of observable matter with a hidden gauge theory.
 Such a framework is the canonical setup today to explain dark matter, although dark matter is considered 
 to be non-relativistic today.
 It is plausible that early interactions may have
 produced acceleration in the observable universe.

In this section we will like to bring this line of thought to its natural 
conclusion. This advocates replacing also four-dimensional gravity by a large N gauge theory.
There are several extra motivations for this leap, some conceptual, others practical:

$\bullet$ Closed string theory generically predicts gravity. Fundamental string theories
provide a consistent (perturbative) quantization of gravity.
Despite its successes, string theory, although  well defined at
energies below or at the string
scale,
breaks down at energies close to the Planck scale. In particular,
the perturbation theory breaks down due to the strong effective
gravitational coupling. Despite speculations \cite{wittat1}-\cite{ooguri}, 
the nature of the
extreme UV degrees of freedom of
the theory is still obscure .
Perturbative closed string theory is essentially a cutoff theory of
gravity and other interactions. This is obvious at the one-loop
level of closed string theory, where the theory has a (smart indeed) cutoff at the string
scale, implemented by Schwinger parameters confined to the
fundamental domain of the torus. A similar structure persists at
higher orders in perturbation theory\footnote{This is not the case in open strings but they do not contain gravity albeit in the form that is advocated later.}.
In perturbative string theory, the string scale is much lower than
the Planck scale. Taking this at face value, we do not expect the theory to give
useful information about physics at energies hierarchically  higher than the
string scale, namely around or above the Planck scale without running at a singularity/strong coupling.

It would seem that non-perturbative dualities might give a way
out, since they provide information about strong coupling physics.
Indeed non-perturbative dualities relate theories with different
(dimensionless) couplings and string scales.
This is however not the case for gravity, since any
non-perturbative duality we know, leaves the Planck scale fixed,
and thus cannot address questions on physics at or beyond the
Planck scale.

$\bullet$ Since the early work of 't Hooft \cite{hoof} it was understood that
the low energy limit of large N-gauge theories is described by some string theory.
The gauge theory versus string theory/gravity
correspondence \cite{mald} is a more precise indication that gravity can be realized as an
effective theory of a four-dimensional gauge theory. The inverse
is also true: fundamental string theory in some backgrounds
describes the physics of theories that at low energy are standard
gauge theories. Although bulk-boundary duality is a concept
transcending that of four-dimensional gauge theories, it is most
powerful in the four-dimensional cases.

The lesson of AdS/CFT correspondence is that any
large-N gauge theory has a dual gravity/string theory\footnote{It happens sometimes that the 
string  has low tension 
and there is no good gravity-alone description.}.
The suggestion of 't Hooft that gravity must be holographic \cite{holog1,holog2}
indicates that a gravity theory must have a dual gauge theory
description.

$\bullet$ A standard gauge theory realization of four-dimensional  gravity generically
predicts massive composite gravitons. The graviton is the spin-two glueball generated out of the vacuum by the
stress-tensor of the theory. Confinement typically comes together with a mass gap.
A graviton mass is severely constrained by observations. Its presence may have two potential advantages.
It predicts an intrinsic  cosmological constant that may be of the
order of magnitude observed today, if the graviton mass modifies
gravity at or beyond the horizon today \cite{report}. Also, the fact that the
graviton is a bound state, allows for a  mechanism
to suppress the observable cosmological constant.
In particular, the graviton {\sl does not} directly couple
to the standard ``vacuum energy" of the SM
fields.

\bu Last but not least, gauge theory so far has been able to explain adequately the Bekenstein-Hawking
entropy of black holes, seeding the idea of bulk-boundary correspondence.

Therefore, the idea is that the building blocks of a theory of
all interactions are four-dimensional  gauge theories. Such theories
are special in many respects both
nature-wise and mathematics-wise. Four-dimensional  gravity is  an
effective, almost classical theory, emanating from a large-N sector 
of the gauge
theory.

The approach advocated here has similarities with ideas in
\cite{sundrum1,sundrum2} and \cite{zee}.
The qualitative model of \cite{sundrum1}, is somewhat different since the SM
particles are not charged under the strong gauge group. Gravity here is
mediated by (heavy) messager matter charged both under the SM
group and the strong gauge group, as suggested by
gauge-theory/string theory correspondence.
In fact, a light scalar graviton can be a meson but not a spin-two
one.
There is also some similarity with the idea of deconstruction
\cite{deconstr}, but here it is gravity rather than higher
dimensional matter theories that is realized by the gauge theory.

There are  direct similarities with attempts to describe
fundamental string theory in terms of matrix models
\cite{c=11}-\cite{matrix2}. Here, however, the gauge physics is four
dimensional and provides a wider class of gravity theories.
Moreover, a four-dimensional large N gauge theory, although more
complicated than a standard Matrix model gives a better intuitive
handle on the physics.

Consider a large N-gauge theory with gauge group $\gn$ and large-N matter (scalars and
fermions) that we
will not specify at the moment.
We would like the UV theory to be  conformal\footnote{It might be possible 
to also realize this
with asymptotically free theories. The bulk-boundary correspondence 
is not, however, understood today in such cases.},
so that it is a well-defined theory at all scales. This
will put constraints on the allowed types of large-N matter content.

At low energy, the effective degrees of freedom are colorless
glueballs as well as mesons (baryons are heavy at large N,
\cite{wittenn}).
Among the effective low-energy degrees of freedom there is always
a spin-two particle (that is generated from the gauge theory vacuum by the
total stress tensor of theory). We will also assume that this theory is confining
in the IR,
and will therefore have a mass gap.
This is not difficult to achieve with today's technology.
The spin-two glueball  will be therefore massive.
However, on general principles (conservation of the gauge theory
stress-tensor) we expect to have a spin-two gauge invariance (that
may be spontaneously broken by the gauge theory vacuum).
Thus, the interactions of this particle, are those of a massive
graviton.

There are, however, other universal composites.
Let us consider for simplicity an $SU(N)$ pure gauge theory.
The leading operators, that are expected to create glueballs
out of the vacuum are a scalar (the ``dilaton")
$\phi\to Tr[F_{\m\n}F^{\m\n}]$, a spin-2 ``graviton" $g_{\m\n}\to
tr[F_{\m\r}{F^{\r}}_{\n}-{1\over 4}\delta_{\m\n}F_{\r\s}F^{\r\s}]$
and a pseudoscalar ``axion" $a\to
\e^{\m\\r\s}Tr[F_{\m\n}F_{\r\s}]$ .
These particles will be massive, and their interactions at low energy are non-perturbative from the
point of view of the gauge theory.
In this relatively simple theory, with two parameters, the mass scale $\Lambda$ and
$N$, the masses are expected to be of order $\Lambda$ and the
interactions controlled at large N by $g_s\sim {1\over
\sqrt{N}}$.
In  particular, three point couplings scale as $1/\sqrt{N}$.

As first advocated by 't Hooft \cite{hoof}, the effective interactions
of the glueballs are expected to be described by an effective
string theory in the low energy regime. Most probably, the
world-sheets of this string theory are discrete, and only a tuning
of the gauge theory (double scaling limit?) might give rise to a
continuous string.

Unlike fundamental string theory, the graviton here is a
bound state of glue, and in the UV, the proper description of its
interactions are in terms of gluons. Thus, in this theory, the low
energy theory is string-like.
But the hard scattering of ``gravitons" is described by
perturbative gauge theory, while their soft scattering by an
effective (massive) gravity/string-theory.
In particular, gravitational interactions turn off at high energy
due to asymptotic freedom.

There are several immediate questions that beg to be answered in such a scheme.

\vskip .7cm

{\bf (A)} The effective graviton must have a mass that is very small (probably
of the order of the inverse horizon size) in order not to be upset by current data.
Just lowering the scale $\Lambda$ of the gauge theory is not
enough. A simple gauge theory with an ultra-low $\Lambda$ (of the order of the inverse horizon size today), has
light gravitons that are on the other hand very loosely bound states
with a size comparable to that of the universe.
We need that their size is hierarchically larger than
their mass.
An important issue is whether a small mass for the graviton is
technically natural. It is conceivable that coordinate invariance
protects the graviton mass as gauge invariance does for the photon
mass.

Moreover, the other generic low lying scalars (dilaton and axion as
well as the spin-0 component of the graviton) must be substantially
heavier so that we are not again upset by data.

What types of large N-gauge theories have a
small or no mass gap? What determines the mass gap?
What determines the hierarchy of
masses of $\phi,g_{\mu\nu}$ and $a$?
Although, there has been considerable efforts to answer such
questions for several gauge theories, no unifying picture exists yet.
This is due to the fact that these questions involve
non-perturbative gauge theory physics. It is also due to the fact that glueballs
have been conspicuously absent from particle physics experiments.
Such questions may be studied using the general ideas of AdS/CFT
correspondence and its generalizations.

An important lesson from AdS/CFT correspondence is that the
gravity dual to four-dimensional gauge theory is five-dimensional
(with additional compact dimensions if extra (adjoint) scalar matter appears
in the gauge theory).
Polyakov has advocated general reasons why this is expected
\cite{polyakov}.
Indeed counting the degrees of freedom of massive
$g_{\mu\nu},\phi,a$ we could expect that their effective
interaction can be described by a five-dimensional massless graviton
as well as five-dimensional scalars $\phi$ and $a$.
The non-trivial gauge theory vacuum should correspond to a
nontrivial background of the five-dimensional theory (as
$AdS_5\times S^5$ describes N=4 super Yang-Mills via AdS/CFT
duality). UV scale invariance 
implies an $AdS_5$ asymptotic region in the effective space-time.

\vskip .7cm

 {\bf (B)} At low energy in the gauge theory
(if it is confining), the effective physics is described by some
string theory (at large N). Also non-confining
theories have a string description as AdS/CFT indicates but only
for the gauge singlet sector.
The important question is: what are the scales of  the string theory/gravity in terms
of the fundamental scales of the gauge theory?
The AdS/CFT paradigm is suggestive.
 Here, on the string theory side there
are three parameters: The AdS radius $R$, the string scale $l_s$
and the string coupling $1/N$. On the gauge theory side there are
only two: N and the 't Hooft coupling $\lambda=g^2N$. N=4 super
Yang-Mills is scale invariant in the symmetric vacuum. This
implies that only the ratio $R/l_s$ is observable:
$R/l_s=\l^{1/4}$.

When a mass gap $\Lambda$ is generated because of temperature
effects, it corresponds to a long distance (far away from
boundary) cutoff in AdS, namely the position $r_0$ of the horizon of the AdS
black-hole,
$r_0=\Lambda~ R^2$. The energy on the AdS side is given by
$E=r/R^2$. The cutoff implies a non-trivial effective string
length for the gauge theory, obtained by red-shifting the AdS
string scale at the horizon \cite{polchinski}:
\be
l_s^{\rm eff}\sim l_s{r_0\over
R}\sim {1\over \l^{1/4}\L}
\ee

Finally there could be masses and/or Yukawa couplings in the
large-N gauge theory. They modify the higher-dimensional geometry
by turning on fluxes \cite{non-ads1,non-ads2}

\vskip .7cm

{\bf (C)} Another important question is: how is the SM accommodated in such a
picture?  The expectation is that the SM gauge group is a separate factor from
 the large-N group. It may be  so by fiat, or it may be connected to
 the  large-N gauge group $G_N$ by symmetry breaking. It could
 also be enlarged to a unified group (SU(5), SO(10) etc).
 The standard model particles are neutral under  $\gn$.
 In order for the effective gravity to be felt by the SM fields
 there should be new massive particles charged under both the
 $\gn$ and the SM gauge group.
 Integrating out these particles, the gravitational interaction is
 generated for the standard model particles. Thus, such particles
 are messagers of the gravitational interaction.
 This is analogous to the picture we have of probe D-branes in
 AdS/CFT \cite{probe1}-\cite{probe4} and the messager particles are the analogues
 of the fluctuations of strings stretched between the main set of branes and the SM
 (probe) branes. 

One could also advocate a certain ``unification" in this context:
The theory starts from a simple large-N gauge group which is
broken to a large-N subgroup generating gravity, as well as
``splinters" (the SM or the conventional unified group as well as other hidden sectors). The
massive states communicate gravity to the SM particles.

\vskip .7cm

{\bf (D)} The issue of the cosmological constant is qualitatively
different here. The standard matter loop diagrams that contribute
to the cosmological constant do not couple to gravity here.
Matter loops induce a potential for the graviton. Since the
graviton is composite, its form factors cut-off the matter
contributions at much lower energies (hopefully at $10^{-3}$ eV)
than the matter theory cutoff.
This is similar to the mechanism advocated in \cite{sundrum1}.

\vskip .7cm

 {\bf (E)} As we have learned from AdS/CFT, and expected on more general principles \cite{polyakov}
 the low energy gravitational theory of a large N-gauge theory
 have at least five non-compact dimensions. The obvious  question is:  how is this compatible with the observed
 4-d gravity. Here there are  two complementary ideas that have the right ingredients to turn observable gravity four-dimensional: RS localization and
 brane induced gravity. The RS idea  can be implemented if the
 ''vacuum" of the gauge theory imposes an  effective UV cutoff at the position of
 the SM branes. This ensures that gravity is four-dimensional in the IR.
  Brane induced gravity (BIG)is always present,  and it ensures gravity is four-dimensional 
  in the UV. Since it comes from the loops of the SM fields, the transition scale associated to it
  is $M_{BIG}\geq M_{Z}$ \cite{ktt2,report}. There will be no constraints if $\ell_{AdS}\sim M_{BIG}\sim 1$ TeV
  where $\ell_{AdS}$ is measured at the SM branes' position.

\vskip .7cm

{\bf (F)} Some of these questions can be put in perspective by
utilizing the essentials of the D-brane picture which underlies gauge-theory/gravity correspondence.
They provide a close link between gauge theory and gravity.

The large N gauge group $\gn$ can be represented as a heavy (large
N) black brane. The SM gauge group can be viewed as some collection
of a few probe branes in the background of the black-hole.
Ideally,
integrating out the strings that connect the probe branes with the
central stack (massive matter charged under both $G_N$ and SM)
induce effective gravitational interactions for the SM fields.

\vskip .7cm
{\bf (H)} The approach described here has a potentially serious problem:
it relies on non-perturbative physics. Typically such a problem
proves fatal. However here we would like to advocate a
5-dimensional gravitational approach to the problem.
Several of the questions described above can be attacked in this
fashion, namely determining the 5-d action and its vacuum solution
and tuning it to achieve small graviton mass and correct
gravitational interactions for standard model particles.
In the next subsections we start a preliminary investigation of
some simple issues in this context. Whether a fully workable model
can emerge remains to be seen.

\section{Acknowledgments}

I would like to thank C. Defayet, A. Kehagias, I, Papadimitriou, C. Sfetsos, K.
Stelle, N. Tetradis, R. Woodard nd especially  K. Skenderis, for fruitful discussions.

This work was partially supported by
INTAS grant, 03-51-6346, RTN contracts MRTN-CT-2004-005104 and
MRTN-CT-2004-503369,  CNRS PICS 2530 and 3059, and by a European Union Excellence Grant,
MEXT-CT-2003-509661.

\appendix

\vskip 10mm
 \renewcommand{\theequation}{\thesection.\arabic{equation}}
\centerline{\Large\bf Appendices}

\addcontentsline{toc}{section}{Appendices}

\section{Conventions and useful  formulae\label{apa}}
\def\G{\Gamma}

We use (-+++) signature for the space-time metric.

The Riemann tensor is:
\be
{R^{\l}}_{\m\n\r}=\p_{\n}{\G_{\m\r}}^{\l}-\p_{\r}{\G_{\m\n}}^{\l}
-{\G_{\m\n}}^{\s}{\G_{\s\r}}^{\l}+{\G_{\m\r}}^{\s}{\G_{\s\n}}^{\l}
\label{44}\ee
while the  Ricci tensor is
\be
R_{\m\n}={R^{\l}}_{\m\l\n}
\label{54}\ee
With these conventions a sphere has constant positive curvature.

Some useful Bianchi identities are
\be
\nabla_{\m}R_{\n\a}-\nabla_{\n}R_{\m\a}=-\nabla_{\r}{R_{\m\n;\a}}^{\r}
\ee
from which it follows that
\be
\nabla^{\m}R_{\m\n}={1\over 2}\nabla_{\n}R
\ee

We will now present the variations of  $R^2$ effective actions

\be
Z^{(0)}_{\m\n}\equiv {1\over \sqrt{-g}}{\delta\over \delta g^{\m\n}}\int d^4x \sqrt{-g}~
R_{\m\n;\r\s}R^{\m\n;\r\s} = 2R_{\m\r;\s\tau}{R_{\n}}^{\r;\s\tau}-{1\over
2}g_{\m\n}R_{\a\b;\r\s}R^{\a\b;\r\s}-\ee
$$
-4\square
R_{\m\n}+2\nabla_{\m}\nabla_{\n}R-4R_{\m\r}{R^{\r}}_{\n}+4R^{\r\s}R_{\r\m;\s\n}
$$

\be
Z^{(1)}_{\m\n}={1\over \sqrt{-g}}{\delta\over \delta g^{\m\n}}\int d^4x \sqrt{-g}~ R^2
=
-2\nabla_{\m}\nabla_{\n}R+2g_{\m\n}\square R-{1\over
2}g_{\m\n}R^2+2RR_{\m\n} \ee

\be
Z^{(2)}_{\m\n} ={1\over \sqrt{-g}}{\delta\over \delta g^{\m\n}}\int d^4x
\sqrt{-g}~R_{\m\n}R^{\m\n}=-2\nabla^{\rho}\nabla_{\n}R_{\r\m}+
\ee
$$
-2\nabla^{\rho}\nabla_{\n}R_{\r\m}+\square R_{\m\n}+{1\over
2}g_{\m\n}\square R+2{R_{\m}}^{\r}R_{\r\n}-{1\over
2}g_{\m\n}R^{\r\s}R_{\r\s}
$$
We have the following relation
\be
Z^{(0)}_{\m\n}=-Z^{(1)}_{\m\n}+4Z^{(2)}_{\m\n} \label{1} \ee
and for conformally
flat space-times
\be Z^{(2)}_{\m\n}={1\over 3}Z^{(1)}_{\m\n} \label{2}\ee

Consider a RW background, 
\be
ds^2=\gamma_{\m\n}dx^{\m}dx^{\n}=-dt^2+a^2 \zeta_{ij}dx^idx^j
\ee
with $\zeta_{ij}$ a maximally symmetric metric with constant curvature.

We use also
\be {\dot a\over a}=H\sp {\ddot a\over a}=\dot H+H^2\sp
{a^{(3)}\over a}=\ddot H+3H\dot H+H^3
\ee
\be
{a^{(4)}\over a}=H^{(3)}+4H\ddot
H+3\dot H^2+6H^2\dot H+H^4 \ee

For the Einstein tensor $G_{\m\n}=R_{\m\n}-{1\over 2}\gamma_{\m\n}R$ we
obtain

\be G_{00}=3\left[H^2+{k\over a^2}\right]\sp G_{ij}=-
\left(2\dot H+3H^2+{k\over a^2}\right)a^2\zeta_{ij} \ee
while the other tensors become
\be Z^{(1)}_{00}=18\left[2{\dot a \over
a}{a^{(3)}\over a}    -{\ddot a^2\over a^2}+2{\dot a^2\over a^2}{\ddot
a\over a}-3{\dot a^4\over a^4}-2{k\over a^2}{\dot a^2\over a^2}+{k^2\over
a^4}\right]= \ee
$$
=18\left[2H\ddot H-\dot H^2+6H^2\dot H-2{k\over a^2}H^2+{k^2\over
a^4}\right]
$$

\be Z^{(1)}_{ij}=6\left[-2{a^{(4)}\over a}-4{\dot a \over a}{a^{(3)}\over
a}    -3{\ddot a^2\over a^2}+12{\dot a^2\over a^2}{\ddot a\over a}-3{\dot
a^4\over a^4}-2{k\over a^2}\left({\dot a^2\over a^2}-2{\ddot a\over
a}\right)+{k^2\over a^4}\right]a^2 \zeta_{ij} \ee
$$
=-6\left[2H^{(3)} +12 H\ddot H+9\dot H^2 +18H^2\dot H -2{k\over a^2}(2\dot
H+H^2)-{k^2\over a^4}\right]a^2 \zeta_{ij}
$$

The values of $Z^{(0)}$ and $Z^{(2)}$ can be calculated from (\ref{1}),
(\ref{2}).

Since
\be
Z_{\m\n}=B' \left(Z^{(2)}_{\m\n}-{1\over 3}Z^{(1)}_{\m\n}\right)-{B\over
3}Z^{(1)}_{\m\n}
\ee
this boils to
\be
Z_{\m\n}=-{B\over
3}Z^{(1)}_{\m\n}
\ee
 for a RW background.

We also have
\be
\square R=-6\left[{a^{(4)}\over a}+6{\dot a a^{(3)}\over a^2}+{\ddot a^2\over a^2}-2{\dot a^2\ddot a\over a^3}-6{\dot a^4\over a^4}\right]
+12{k\over a^2}\left[{\ddot a\over a}+3{\dot a^2\over a^2}\right]\ee
$$
=-6(12H^2\dot H+4\dot H^2+7H\ddot H+H^{(3)})+12{k\over a^2}(\dot
H+4H^2)
$$
The Gauss-Bonnet invariant for an FRW universe is
\be
G=24{\ddot a \over a}\left[{\dot a^2\over a^2}+{k\over a^2}\right]= 24(\dot H+H^2)\left[H^2+{k\over a^2}\right]
\ee
while the scalar curvature is
\be
 R=-6\left({\ddot a \over a}+{\dot a^2\over
a^2}+{k\over a^2}\right)=-6\left(\dot H+2H^2+{k\over a^2}\right)
\ee

\section{Conformal Anomalies in Four Dimensions\label{anomaly}}

Four-dimensional conformal field theories, defined on curved
backgrounds suffer from conformal anomalies (see  \cite{duff},
\cite{bd} for a review and references) They describe the breaking
of conformal invariance due to quantum effects (regularization and
renormalization).

The general form of the anomaly, for a classically conformally
invariant theory, is given by the expectation value of the trace
of the stress tensor\footnote{We assume here that the 
values of other background fields except the metric are zero.}. The general formula is

\be <{T^{\m}}_{\m}>=-a~ C^2-c~G+b~  \square R \label{conf}\ee

where $C_{\m\n;\r\s}$ is the Weyl tensor \be
C_{\m\n;\r\s}=R_{\m\n;\r\s}-{1\over
2}(g_{\m\r}R_{\s\n}-g_{\m\s}R_{\r\n}-g_{\n\r}R_{\s\m}+g_{\n\s}R_{\r\m})
+{1\over 6}(g_{\m\r}g_{\s\n}-g_{\m\s}g_{\r\n})R \ee so that \be
C^{\m\n;\r\s}C_{\m\n;\r\s}=R^{\m\n;\r\s}R_{\m\n;\r\s}-2R^{\m\n}R_{\m\n}+{1\over
3}R^2 \ee and \be G=R_{\m\n\r\s}R^{\m\n\r\s}-4R_{\m\n}R^{\m\n}+R^2 \ee is
the Gauss-Bonnet density. The coefficients $a,c$ are scheme-independent
while $b$ is scheme-depended. This is as well, since it can be changed by
adding the square of the scalar curvature in the action. Under the Weyl
transformation $\delta g_{\m\n}=\phi ~g_{\m\n}$ the effective action
changes by definition by

\be \delta S=\int d^4x\sqrt{-g}~ \phi ~ {T^{\m}}_{\m} \ee

Since under a Weyl transformation

\be \delta\int d^4x \sqrt{-g} ~R^2 =-6\int d^4x\sqrt{-g}~ \phi~
\square R \ee
we can remove the scheme-dependent contribution by
adding

\be {b\over 6} \int d^4x \sqrt{-g} ~R^2 \ee to the effective
action.

To analyse the individual contributions of free massless fields
 we may
parameterize the trace as
 \be <{T^{\m}}_{\m}>=-{\a C^2+\b(R^{\r\s}R_{\r\s}-{1\over
3}R^2)-\eta\square R-\delta R^2\over 180(4\pi)^2}\label{conf2}\ee

Then, in dimensional regularization, the contributions from fields transforming in the
representation (A,B) of the $SU(2)\times SL(2)$ subgroup of the Poincar\'e
group is given in the following table \cite{bd}

\bigskip
\begin{center}
\begin{tabular}{|c|c|c|c|c|}
  \hline
  {(A,B)} & $\a$ & $\b$ & $\eta$ & $\d$ \\
  \hline
  &&&& \\
  (0,0) & -1 & -1 & 6-30$\xi$ & 90$\left(\xi-{1\over 6}\right)^2$ \\
    &&&& \\
  $\left({1\over 2},0\right)\oplus \left(0,{1\over 2}\right)$ & -${7\over 2}$ & -11 & 6 & 0 \\
    &&&& \\
  $\left({1\over 2},{1\over 2}\right)$ & 11 & -64 & -6 & 5 \\
    &&&& \\
  (1,0) & -33 & 27 & 12 & ${5\over 2}$ \\
    &&&& \\
   $\left({1\over 2},1\right)$ & ${291\over 4}$ & X & X & -${61\over 8}$ \\
     &&&& \\
  (1,1) & -189 & X & X & ${747\over 4}$ \\
  \hline
\end{tabular}
\end{center}
where $\xi$ is the conformal coupling of scalars. X stands for
terms that are forbidden by consistency conditions. The
contributions above are off-shell. For example, to get the
contribution of the photon we must subtract that of two scalars
from the vector. For physical fields, in dimensional
regularization we obtain

\bigskip
\begin{center}
\begin{tabular}{|c|c|c|c|c|c|}
  \hline
  {\rm Spin} & (A,B)& $\a$ & $\b$ & $\eta$ & $\d$ \\
  \hline
  &&&&& \\
  0&(0,0) & -1 & -1 & 6-30$\xi$ & 90$\left(\xi-{1\over 6}\right)^2$ \\
    &&&&& \\
  ${1\over 2}$&$\left({1\over 2},0\right)\oplus \left(0,{1\over 2}\right)$ & -${7\over 2}$ & -11 & 6 & 0 \\
   & &&&& \\
  1&$\left({1\over 2},{1\over 2}\right)-2(0,0)$ & 13 & -62 & -18 & 0 \\
    &&&&& \\
   ${3\over 2}$&$\left({1\over 2},1\right)-2\left({1\over 2}
   ,0\right)$ & ${233\over 4}$ & X & X & -${61\over 8}$ \\
    & &&&& \\
  2&$(1,1)+(0,0)-2\left({1\over 2},{1\over 2}\right)$ & -212 & X & X & ${717\over 4}$ \\
  \hline
\end{tabular}
\end{center}

We can thus obtain for the scheme independent coefficients of a
classically conformal field theory ($\d=0$ for (\ref{conf}) to be
compatible with (\ref{conf2}))
as

\be 
c={1\over 360(4\pi)^2}(N_{\rm scalar}+11N_{\rm
fermion}+62N_{\rm vector}) 
\ee 
where the fermions are Dirac and
\be 
a=-{1\over 120(4\pi)^2}(N_{\rm scalar}+6N_{\rm
fermion}+12N_{\rm vector})\label{conf20} 
\ee

Around the free conformal point, the scalars must be conformally coupled
($\xi=1/6$), and the fermions and vectors massless.

\section{Adding a Gauss-Bonnet term in the bulk\label{char}}

We will present  here an example of the effect of higher derivative terms in the bulk equations.
Stringy corrections are expected to give such higher derivative terms in bulk. We choose here 
the simplest one, namely the Gauss-Bonnet term for purposes of illustration.

We start from
\be
S=\int d^{5}x~ \sqrt{-g} \left( M^3 R -\Lambda_5 +\zeta~ G\right)
+\int d^{4} x\sqrt{-\hat g} \,\left( -V+{\cal L}_b^{mat} \right),
\label{0010}
\ee
According to \cite{Charmousis} the FRW equation that asymptotes
properly to the $\zeta\to 0$ case is (we consider flat 3-space), 
\be
\left({\rho+V\over 32\zeta}\right)^2=\left(H^2+U
\right)^3+C\left(H^2+U\right)^2+{C^2\over 4}\left(H^2+U\right)
\ee
where
\be
C={3\over 4\zeta}\sqrt{1+{2\over 3}\zeta~ {\Lambda_5\over M^6}+{8\zeta}\chi}\sp U={1\over 4\zeta}-{3\over 16}C
\sp \chi={\mu^4\over a^4}\ee
All dimensionfull parameters ($\zeta$, $V$, $\Lambda_5$) are
measured in units of $M$.

Expanding for small $\zeta$ we obtain
\be
{(\rho+V)^2\over 12}=12H^2-\Lambda_5-12\chi+{\zeta\over 18}\left[
2.24^2H^4-48H^2(\Lambda_5+12\chi)-(\Lambda_5+12\chi)^2\right]+{\cal
O}(\zeta^2)
\ee
which can be inverted to
\be
H^2={12\chi+\Lambda_5\over 12}+{(\rho+V)^2\over 12^2}-{2\over
3}\zeta\left[3{(12\chi+\Lambda_5)^2\over 12^2}+{(\rho+V)^2\over
12}{(12\chi+\Lambda_5)\over 12}+8{(\rho+V)^4\over 12^4}\right]+{\cal
O}(\zeta^2)
\ee
We now reinsert $M$

\be
H^2={\chi\over M^2}+{\Lambda_5\over 12M^3}+{(\rho+V)^2\over 12^2M^6}-{2\over
3}{\zeta\over M}\left[{3\over M^2}\left({\chi\over M^2}+{\Lambda_5\over 12M^3}\right)^2+{(\rho+V)^2\over
12 M^8}\left({\chi\over M^2}+{\Lambda_5\over 12M^3}\right)+
\right.\ee
$$\left.+
8{(\rho+V)^4\over 12^4M^{14}}\right]+{\cal
O}(\zeta^2)
$$
\def\chir{\chi_{\rm rad}}

The RS fine-tuning that cancels the effective cosmological
constant now becomes
\be
\Lambda_5=-{V^2\over 12M^3}-8\zeta{V^4\over 12^4 M^{12}}+{\cal
O}(\zeta^2)
\ee

Redefining now $\chi={V\over 72 M^4}\chir$ the previous equation
becomes

\be
H^2={\rho^2\over 144M^6}+{V\over 72 M^6}(\rho+\chir)-{2\over 3}{\zeta
\over M}{V^4\over 12^4 M^{14}}\left[3\left(2{\chir\over V}-1\right)^2+
\right.\ee
$$
+\left.
12\left(2{\chir\over V}-1\right)\left({\rho\over V}+1\right)^2
+8\left({\rho\over V}+1\right)^4+1\right]+{\cal O}(\zeta^2)$$

where $\chir\sim {1\over a^4}$.

Notice that the extra terms obtained on and above the terms of RS , are of the form $\chir\rho$ and 
$\chir^2$, as in the expansion at quadratic order of the SCGT solution (\ref{sol}).

\section{\label{D}The general case with $D\not= 0$}

We will consider here the contribution of the four-dimensional graviton to the conformal anomaly.
 Apart from modifying c to order $1/N^2$ it will also give a non-zero coefficient $D$.
 
 A discussion is in order here, concerning the possible signs of $D$.
 As can be seen in the tables of appendix \ref{anomaly}
 from the standard massless particles, it is only the gravitino that contributes negatively
to $D$. However, at least 24 gravitini are needed to offset the graviton contribution.
Thus, we may assume that $D>0$. However, it should be kept in mind that the possibility that
$D<0$ is open. For example there could be  higher spin particles as in string theory which could induce a negative $D$.

The system of three equations (\ref{frwg},\ref{chia},\ref{rho}) after rescaling to dimensionless 
variables (\ref{scaling}) and adding the $R^2$ term, are
\be
4\left[H^2+{k\over a^2}\right]-4\left[H^2+{k\over a^2}\right]^2=\rho+\chi
\label{frwg2}\ee
\be
\dot\chi+4H\chi=T+H\left[X-36D\left(\dot H+2H^2+{k\over a^2}\right)^2\right]
\label{chi2}\ee
\be
\dot \rho+3H(\rho+p)=-T
\label{rho2}\ee
These can be 
massaged further.
Differentiating (\ref{frwg2}) we can obtain $\dot H$.

\be
\dot H+2H^2+{k\over a^2}={\dot \rho+\dot \chi\over
8H\left[(1-2\left(H^2+{k\over a^2}\right)\right]}+2\left(H^2+{k\over a^2}\right)
\label{e4}\ee
$$
=
{\dot \rho+\dot \chi+8H(\rho+\chi)-16H\left(H^2+{k\over a^2}\right)\over 
8H\left[(1-2\left(H^2+{k\over a^2}\right)\right]}
$$

from which we obtain
\be
(1-\xi)\left(\dot H-{k\over a^2}\right)=2\xi\left[H^2+{k\over a^2}\right]-1+
\ee
$$+
\e\sqrt{1-4\left[H^2+{k\over a^2}\right]+4\xi\left[H^2+{k\over a^2}\right]^2+(1-\xi)(\rho+X-3p)}
$$
where $\xi=1-9D$.
If D is generated solely by the graviton then 
\be
D={717\over 360(4\pi)^2}\sp \xi\simeq 0.87
\ee
although other values are also possible.

For the square root to be real we must have
\be
(i) ~~~~~~~~ H^2+{k\over a^2}\leq {1-\sqrt{1-\xi}\sqrt{1-\xi(\rho+X-3p)}\over 2\xi}\sp \e=1
\ee
\be
(ii) ~~~~~~~~ H^2+{k\over a^2}\geq {1+\sqrt{1-\xi}\sqrt{1-\xi(\rho+X-3p)}\over 2\xi}\sp \e=-1
\ee

The fixed points satisfy 
\be
H^2+{k\over a^2}={1\pm\sqrt{1-\xi(X+\rho-3p)}\over 2\xi}
\ee
one being in region (i), the other in region (ii).

\subsection{RS with the graviton contribution}

We will now rediscuss the pure RS case, with the graviton contribution included.
Setting $T=X=0$ in the formulae of the previous section
we obtain

\be
4\left[H^2+{k\over a^2}\right]-4\left[H^2+{k\over a^2}\right]^2=\rho+\chi
\label{frwg5}\ee
\be
\dot
\chi+4H\chi=-36DH\left(\dot H+2H^2+{k\over a^2}\right)^2
\label{chi5}\ee
\be
\dot \rho+3H(\rho+p)=0
\label{rho5}\ee

To make things more transparent we will first consider the case where the observable energy density is also
radiation.
When $w=1/3$ we can lump together $\rho+\chi\to \rho$ and the equations become
\be
4\left[H^2+{k\over a^2}\right]-4\left[H^2+{k\over a^2}\right]^2=\rho
\label{frwg4}\ee
\be
\dot
\rho+4H\rho=-36DH\left(\dot H+2H^2+{k\over a^2}\right)^2
\label{chi4}\ee
We can replace (\ref{chi4}) by
\be
9D\left(\dot H-{k\over a^2}\right)^2+2\left(1-2\xi\left[H^2+{k\over a^2}\right]\right)\left(\dot H-{k\over a^2}\right)
+4\left[H^2+{k\over a^2}\right]\left(1-\xi\left[H^2+{k\over a^2}\right]\right)=0
\label{nonl}\ee
where $\xi=1-9D$.
This can be solved as
\be
(1-\xi)\left(\dot H-{k\over a^2}\right)=2\xi\left[H^2+{k\over a^2}\right]-1+\e\sqrt{1-4\left[H^2+{k\over a^2}\right]+4\xi\left[H^2+{k\over a^2}\right]^2}
\ee
For the equation (\ref{nonl}) to have solutions the quantity under the square root must be non-negative.
This gives two possibilities:
\be
(i) ~~~~~~~~ H^2+{k\over a^2}\leq {1-\sqrt{1-\xi}\over 2\xi}\sp (ii) ~~~~~~~~ H^2+{k\over a^2}\geq {1+\sqrt{1-\xi}\over 2\xi}
\ee
In the $\e=1$ branch we are in the region (i).
The only possible fixed point is $H^2+{k\over a^2}=0$ valid  if $k\not= 1$.
Solving the linearized equation around that fixed point we find the standard evolution of a (curved) universe filled with dilute radiation 
\be
a^2\simeq a(0)^2-Ct-kt^2
\ee
The fixed point is attractive.
Note that the $R^2$ correction is irrelevant in the neighborhood of this fixed point.
Moreover, for the solution in region (i) we have always deceleration.

In the $\e=-1$ branch we are in region (ii).
There is a non-trivial fixed point, a deformation of the Starobinsky one, with
\be
H^2+{k\over a^2}={1\over \xi}
\ee
providing an inflating universe
\be
a(t)={\xi\over 2}\left[e^{t\over \sqrt{\xi}}+ke^{-{t\over \sqrt{\xi}}}\right]
\ee
This fixed point is also attractive.
First order perturbation theory for $H=H_*+\delta H$, $\rho=\delta\rho$, $\chi=\chi_*+\delta\chi$
gives
\be
\delta\rho+\delta\chi-8H_*(1-2H_*^2)\delta H=0
\ee
\be
{\partial\over \partial t}(\delta\chi-144DH_*^3 \delta H)+4H_*(\delta\chi-144DH_*^3 \delta H)=0
\ee
\be
\dot {\delta\rho}+3H_*(\delta\rho+\delta p)=0
\ee
and indicates that the fixed point is attractive (stable).

 The whole region (ii) is accelerating.
It should be noted however, that the presence of several non-minimally coupled scalars 
may destroy this fixed point since they will make $9D>1$. However, in this case $X\not =0$
and a more detailed analysis is necessary.


\begin{thebibliography}{999}

\bibitem{report}  E.~Kiritsis,
  ``D-branes in standard model building, gravity and cosmology,''
   Fortsch.\ Phys.\  {\bf 52} (2004) 200 and
to appear in Physics Reports, \hre{hep-th}{0310001}.


  
  \bibitem{oreport}
F.~Quevedo,
``Lectures on string / brane cosmology,''
Class.\ Quant.\ Grav.\  {\bf 19} (2002) 5721
\hre{hep-th}{0210292}.


\bibitem{review2}
R.~Maartens,
``Brane-world gravity,''
\hre{gr-qc}{0312059}.

\bibitem{rep2} 
 E.~Kiritsis,
  ``Brane-bulk energy exchange and cosmological acceleration,''
  Fortsch.\ Phys.\  {\bf 52} (2004) 568
 \hre{hep-th}{0503189}.

\bibitem{review1}
M.~Trodden and S.~M.~Carroll,
``TASI lectures: Introduction to cosmology,''
\hre{astro-ph}{0401547}.


\bibitem{brax}
P.~Brax, C.~van de Bruck and A.~C.~Davis, ``Brane world cosmology,''\\
\hre{hep-th}{0404011}.

\bibitem{csaki}
C.~Csaki,
``TASI lectures on extra dimensions and branes,''
\hre{hep-ph}{0404096}.

\bibitem{Silverstein:2004id}
E.~Silverstein,
``TASI / PiTP / ISS lectures on moduli and microphysics,''\\
\hre{hep-th}{0405068}.

\bibitem{Gabadadze:2004dq}
G.~Gabadadze,
``Looking at the cosmological constant from infinite-volume bulk,''
\hre{hep-th}{0408118}.


\bibitem{oreportf}
U.~H.~Danielsson,
``Lectures on string theory and cosmology,''\\
\hre{hep-th}{0409274}.



\bibitem{rs} L.~Randall and R.~Sundrum,
  ``A large mass hierarchy from a small extra dimension,''
  Phys.\ Rev.\ Lett.\  {\bf 83} (1999) 3370
 \hre{hep-ph}{9905221};\\
   ``An alternative to compactification,''
  Phys.\ Rev.\ Lett.\  {\bf 83} (1999) 4690;\\
 \hre{hep-th}{9906064}.
  
  \bibitem{binetruy}
P.~Bin\'etruy, C.~Deffayet and D.~Langlois,
``Non-conventional cosmology from a brane-universe,''
Nucl. Phys. {\bf B565} (2000) 269\\
\hre{hep-th}{9905012};\\
J.~M.~Cline, C.~Grojean and G.~Servant,
``Cosmological expansion in the presence of extra dimensions,''
Phys.\ Rev.\ Lett.\  {\bf 83} (1999) 4245\\
\hre{hep-ph}{9906523};\\
P.~Bin\'etruy, C.~Deffayet, U.~Ellwanger and D.~Langlois,
``Brane cosmological evolution in a bulk with cosmological constant,''
Phys. Lett. {\bf B477} (2000) 285;\\
\hre{hep-th}{9910219}.


\bibitem{dt}
G.~R.~Dvali and S.~H.~H.~Tye,
``Brane inflation,''
Phys.\ Lett.\ B {\bf 450} (1999) 72
\hre{hep-ph}{9812483}.


\bibitem{mirage}
A.~Kehagias and E.~Kiritsis,
 ``Mirage cosmology,''
  JHEP {\bf 9911} (1999) 022;\\
  \hre{hep-th}{9910174}.


\bibitem{vsl}
E.~Kiritsis,
``Supergravity, D-brane probes and thermal super Yang-Mills:  A comparison,''
  JHEP {\bf 9910} (1999) 010
  \hre{hep-th}{9906206};\\
  S.~H.~S.~Alexander,
  ``On the varying speed of light in a brane-induced FRW universe,''
  JHEP {\bf 0011} (2000) 017
 \hre{hep-th}{9912037}.
  
  \bibitem{bb}
   S.~H.~S.~Alexander,
  ``Inflation from D - anti-D brane annihilation,''
  Phys.\ Rev.\ D {\bf 65} (2002) 023507
  \hre{hep-th}{0105032};\\
C.~P.~Burgess, M.~Majumdar, D.~Nolte, F.~Quevedo, G.~Rajesh and R.~J.~Zhang,
``The inflationary brane-antibrane universe,''
JHEP {\bf 0107} (2001) 047;\\
\hre{hep-th}{0105204};\\
G.~R.~Dvali, Q.~Shafi and S.~Solganik,
``D-brane inflation,'', \hre{hep-th}{0105203}.

\bibitem{big}
G.~R.~Dvali, G.~Gabadadze and M.~Porrati,
``Metastable gravitons and infinite volume extra dimensions,''
Phys.\ Lett.\ B {\bf 484} (2000) 112;
\hre{hep-th}{0002190};\\
``4D gravity on a brane in 5D Minkowski space,''
Phys. Lett. {\bf B485} (2000) 208;\hre{hep-th}{0005016}.

\bibitem{tachyon}
  G.~W.~Gibbons,
  ``Cosmological evolution of the rolling tachyon,''
  Phys.\ Lett.\ B {\bf 537} (2002) 1
  \hre{hep-th}{0204008}.
  
\bibitem{AADD} N.~Arkani-Hamed, S.~Dimopoulos and G.~R.~Dvali,
  ``The hierarchy problem and new dimensions at a millimeter,''
  Phys.\ Lett.\ B {\bf 429} (1998) 263
 \hre{hep-ph}{9803315};\\
I.~Antoniadis, N.~Arkani-Hamed, S.~Dimopoulos and G.~R.~Dvali,
  ``New dimensions at a millimeter to a Fermi and superstrings at a TeV,''
  Phys.\ Lett.\ B {\bf 436} (1998) 257
 \hre{hep-ph}{9804398};\\
  N.~Arkani-Hamed, S.~Dimopoulos and G.~R.~Dvali,
  ``Phenomenology, astrophysics and cosmology of theories with  sub-millimeter
  dimensions and TeV scale quantum gravity,''
  Phys.\ Rev.\ D {\bf 59} (1999) 086004
  \hre{hep-ph}{9807344}.
  
  
  \bibitem{he1} C.~van de Bruck, M.~Dorca, C.~J.~Martins and M.~Parry,
``Cosmological consequences of the brane/bulk interaction,''
Phys.\ Lett.\ B {\bf 495} (2000) 183
\hre{hep-th}{0009056}.

\bibitem{he2} U.~Ellwanger,
  ``Cosmological evolution in compactified Horava-Witten theory induced by
  matter on the branes,''
  Eur.\ Phys.\ J.\ C {\bf 25} (2002) 157
 \hre{hep-th}{0001126}.

\bibitem{he3} A.~Hebecker and J.~March-Russell,
  ``Randall-Sundrum II cosmology, AdS/CFT, and the bulk black hole,''
  Nucl.\ Phys.\ B {\bf 608} (2001) 375
  \hre{hep-ph}{0103214}.

\bibitem{ktt1} E.~Kiritsis, N.~Tetradis and T.~N.~Tomaras,
  ``Induced brane gravity: Realizations and limitations''
  JHEP {\bf 0108} (2001) 012
  \hre{hep-th}{0106050}.








\bibitem{he4}P.~Brax, C.~van de Bruck and A.~C.~Davis,
``Brane-world cosmology, bulk scalars and perturbations,''
JHEP {\bf 0110} (2001) 026
\hre{hep-th}{0108215};

\bibitem{ktt2} E.~Kiritsis, N.~Tetradis and T.~N.~Tomaras,
  ``Induced gravity on RS branes,''
  JHEP {\bf 0203} (2002) 019
  \hre{hep-th}{0202037}.
  
  
\bibitem{he5}
  D.~Langlois, L.~Sorbo and M.~Rodriguez-Martinez,
  ``Cosmology of a brane radiating gravitons into the extra dimension,''
  Phys.\ Rev.\ Lett.\  {\bf 89} (2002) 171301;\\
  \hre{hep-th}{0206146}.

\bibitem{kkttz}
E.~Kiritsis, G.~Kofinas, N.~Tetradis, T.~N.~Tomaras and V.~Zarikas,
``Cosmological evolution with brane-bulk energy exchange,''
JHEP {\bf 0302} (2003) 035;\\
\hre{hep-th}{0207060}.



\bibitem{tetr}
N.~Tetradis,
  ``Cosmological acceleration from energy influx,''
  Phys.\ Lett.\ B {\bf 569} (2003) 1;
  \hre{hep-th}{0211200}.
  
  \bibitem{allpapers1}
  T.~N.~Tomaras,
  ``Brane-world evolution with brane-bulk energy exchange,'';\\
\href{http://www.slac.stanford.edu/spires/find/hep/www?irn=6018939}{SPIRES entry}
{\it Prepared for 27th Johns Hopkins Workshop on Current Problems in 
Particle Theory: Symmetries and Mysteries of M-Theory, Goteborg, Sweden, 24-
26 Aug 2003}

\bibitem{allpapers2}
  P.~S.~Apostolopoulos and N.~Tetradis,
  ``Brane cosmology with matter in the bulk,''
  Class.\ Quant.\ Grav.\  {\bf 21} (2004) 4781
\hre{hep-th}{0404105};\\
  N.~Tetradis,
  ``Brane cosmology with matter in the bulk. II,''
  Class.\ Quant.\ Grav.\  {\bf 21} (2004) 5221
 \hre{hep-th}{0406183}.

\bibitem{allpapers3}
  F.~K.~Diakonos, E.~N.~Saridakis and N.~Tetradis,
  ``Energy from the bulk through parametric resonance,''
  Phys.\ Lett.\ B {\bf 605} (2005) 1
  \hre{hep-th}{0409025}.
  


\bibitem{allpapers4}
  P.~S.~Apostolopoulos and N.~Tetradis,
  ``Brane cosmological evolution with a general bulk matter configuration,''
  Phys.\ Rev.\ D {\bf 71} (2005) 043506
  \hre{hep-th}{0412246}.

 \bibitem{allpapers5}
 P.~S.~Apostolopoulos, N.~Brouzakis, E.~N.~Saridakis and N.~Tetradis,
  ``Mirage effects on the brane,''
  \hre{hep-th}{0502115}.



  \bibitem{RSr1}
  D.~Langlois and L.~Sorbo,
  ``Bulk gravitons from a cosmological brane,''
  Phys.\ Rev.\ D {\bf 68} (2003) 084006
  \hre{hep-th}{0306281}.
  
\bibitem{RSr2}
  E.~Leeper, R.~Maartens and C.~F.~Sopuerta,
  ``Dynamics of radiating braneworlds,''
  Class.\ Quant.\ Grav.\  {\bf 21} (2004) 1125
\hre{gr-qc}{0309080};\\
L.~A.~Gergely, E.~Leeper and R.~Maartens,
  ``Asymmetric radiating brane-world,''
  Phys.\ Rev.\ D {\bf 70} (2004) 104025
\hre{gr-qc}{0408084}.
  
  \bibitem{rsh}
  T.~Shiromizu and D.~Ida,
  ``Anti-de Sitter no hair, AdS/CFT and the brane-world,''
  Phys.\ Rev.\ D {\bf 64} (2001) 044015
\hre{hep-th}{0102035}.
  
  
  \bibitem{mald}
J.~M.~Maldacena,
``The large N limit of superconformal field theories and supergravity,''
Adv.\ Theor.\ Math.\ Phys.\  {\bf 2} (1998) 231
[Int.\ J.\ Theor.\ Phys.\  {\bf 38} (1999) 1113]
\hre{hep-th}{9711200}.

\bibitem{review} 
  O.~Aharony, S.~S.~Gubser, J.~M.~Maldacena, H.~Ooguri and Y.~Oz,
  ``Large N field theories, string theory and gravity,''
  Phys.\ Rept.\  {\bf 323} (2000) 183
  \hre{hep-th}{9905111}.

\bibitem{g0}
J. Maldacena, 1999, unpublished.

\bibitem{g00}
H.~Verlinde,
``Holography and compactification,''
Nucl.\ Phys.\ B {\bf 580} (2000) 264
\hre{hep-th}{9906182};\\
Talk at the ITP Santa Barbara conference "New dimensions in field theory
and string theory"
\href{http://www.kitp.ucsb.edu/online/susy_c99/verlinde/}{http://www.kitp.ucsb.edu/online/susy\_c99/verlinde/}

\bibitem{g000}
E. Witten Remarks at the ITP Santa Barbara conference "New dimensions in field theory
and string theory"
\href{http://www.kitp.ucsb.edu/online/susy_c99/discussion/}
{http://www.kitp.ucsb.edu/online/susy\_c99/discussion/}

\bibitem{g1}
S.~S.~Gubser,
``AdS/CFT and gravity,''
Phys.\ Rev.\ D {\bf 63} (2001) 084017\\
\hre{hep-th}{9912001}.

\bibitem{hawk1}
S.~W.~Hawking, T.~Hertog and H.~S.~Reall,
``Brane new world,''
Phys.\ Rev.\ D {\bf 62} (2000) 043501
 \hre{hep-th}{0003052}.

\bibitem{hawk2}
S.~W.~Hawking, T.~Hertog and H.~S.~Reall,
``Trace anomaly driven inflation,''
Phys.\ Rev.\ D {\bf 63} (2001) 083504 ;
\hre{hep-th}{0010232}.


\bibitem{skenderis}
  S.~de Haro, K.~Skenderis and S.~N.~Solodukhin,
  ``Gravity in warped compactifications and the holographic stress tensor,''
  Class.\ Quant.\ Grav.\  {\bf 18} (2001) 3171;\\
  \hre{hep-th}{0011230}.
  
 


\bibitem{g2}
N.~Arkani-Hamed, M.~Porrati and L.~Randall,
``Holography and phenomenology,''
JHEP {\bf 0108} (2001) 017
\hre{hep-th}{0012148}.

\bibitem{g3}
A.~Hebecker and J.~March-Russell,
``Randall-Sundrum II cosmology, AdS/CFT, and the bulk black hole,''
Nucl.\ Phys.\ B {\bf 608} (2001) 375
\hre{hep-ph}{0103214}.

  \bibitem{star1}
A.~A.~Starobinsky,
``A New Type Of Isotropic Cosmological Models Without Singularity,''
\hspi{752932}{Phys.\ Lett.\ B {\bf 91} (1980) 99.}

\bibitem{Vilenkin}A. Vilenkin, ``Classical And Quantum Cosmology Of The Starobinsky Inflationary Model,''
\hspi{ 1333240}{Phys. Rev. {\bf D32}(1985) 2511.}

\bibitem{star3} A. M. Pelinson, I. L. Shapiro and F. I. Takakura,
``On the stability of the anomaly-induced inflation,''
  Nucl.\ Phys.\ B {\bf 648} (2003) 417
  \hre{hep-ph}{0208184}.
  
  
  
\bibitem{ren1}
E.~Witten,
``Anti-de Sitter space and holography,''
Adv.\ Theor.\ Math.\ Phys.\  {\bf 2} (1998) 253
\hre{hep-th}{9802150}.

\bibitem{ren2}
H.~Liu and A.~A.~Tseytlin,
``D = 4 super Yang-Mills, D = 5 gauged supergravity, and D = 4 conformal
supergravity,''
Nucl.\ Phys.\ B {\bf 533} (1998) 88
\hre{hep-th}{9804083}.

\bibitem{ren3}
M.~Henningson and K.~Skenderis,
``The holographic Weyl anomaly,''
JHEP {\bf 9807} (1998) 023
\hre{hep-th}{9806087};\\
``Holography and the Weyl anomaly,''
Fortsch.\ Phys.\  {\bf 48} (2000) 125\\
\hre{hep-th}{9812032}.

\bibitem{ren4}
V.~Balasubramanian and P.~Kraus,
``A stress tensor for anti-de Sitter gravity,''
Commun.\ Math.\ Phys.\  {\bf 208} (1999) 413
\hre{hep-th}{9902121}.

\bibitem{ren5}
R.~Emparan, C.~V.~Johnson and R.~C.~Myers,
``Surface terms as counterterms in the AdS/CFT correspondence,''
Phys.\ Rev.\ D {\bf 60} (1999) 104001
\hre{hep-th}{9903238}.

\bibitem{ren6}
P.~Kraus, F.~Larsen and R.~Siebelink,
``The gravitational action in asymptotically AdS and flat spacetimes,''
Nucl.\ Phys.\ B {\bf 563} (1999) 259
\hre{hep-th}{9906127}.

\bibitem{ren7}
K.~Skenderis and S.~N.~Solodukhin,
``Quantum effective action from the AdS/CFT correspondence,''
Phys.\ Lett.\ B {\bf 472} (2000) 316
\hre{hep-th}{9910023}.

\bibitem{ren10}
S.~de Haro, S.~N.~Solodukhin and K.~Skenderis,
``Holographic reconstruction of spacetime and renormalization in the  AdS/CFT
correspondence,''
Commun.\ Math.\ Phys.\  {\bf 217} (2001) 595
\hre{hep-th}{0002230}.

\bibitem{ren8}
M.~Bianchi, D.~Z.~Freedman and K.~Skenderis,
``How to go with an RG flow,''
JHEP {\bf 0108} (2001) 041
\hre{hep-th}{0105276};
``Holographic renormalization,''
Nucl.\ Phys.\ B {\bf 631} (2002) 159
\hre{hep-th}{0112119}.

\bibitem{ren9} 
  I.~Papadimitriou and K.~Skenderis,
  ``AdS/CFT correspondence and geometry,''
  \hre{hep-th}{0404176};\\
I.~Papadimitriou and K.~Skenderis,
  ``Correlation functions in holographic RG flows,''
  JHEP {\bf 0410} (2004) 075
  \hre{hep-th}{0407071}.

\bibitem{FG} C. Fefferman and C. Robin Graham, 
``Conformal Invariants", in Elie Cartan et les Mathematiques 
d'aujourd'hui (Asterisque 1985) 95.

\bibitem{GB}
G.~W.~Gibbons and S.~W.~Hawking,
``Action Integrals And Partition Functions In Quantum Gravity,''
\hspi{154130}{Phys.\ Rev.\ D {\bf 15} (1977) 2752}.


\bibitem{cb}
  D.~Z.~Freedman, S.~S.~Gubser, K.~Pilch and N.~P.~Warner,
  ``Continuous distributions of D3-branes and gauged supergravity,''
  JHEP {\bf 0007} (2000) 038;\\
  \hre{hep-th}{9906194}.
  
\bibitem{cb1}
  A.~Brandhuber and K.~Sfetsos,
  ``Wilson loops from multicentre and rotating branes, mass gaps and phase
  structure in gauge theories,''
  Adv.\ Theor.\ Math.\ Phys.\  {\bf 3} (1999) 851
  \hre{hep-th}{9906201}.
  
\bibitem{gppz}
  L.~Girardello, M.~Petrini, M.~Porrati and A.~Zaffaroni,
  ``The supergravity dual of N = 1 super Yang-Mills theory,''
  Nucl.\ Phys.\ B {\bf 569} (2000) 451
  \hre{hep-th}{9909047}.
  
  \bibitem{wittat1}
J.~J.~Atick and E.~Witten,
``The Hagedorn Transition And The Number Of Degrees Of Freedom Of String Theory,''
\hspi{1835610}{Nucl.\ Phys.\ B {\bf 310} (1988) 291}.

\bibitem{wittat2}
E.~Witten,
``Space-Time And Topological Orbifolds,''
\hspi{1877569}{Phys.\ Rev.\ Lett.\  {\bf 61} (1988) 670}.




\bibitem{veneziano}
D.~Amati, M.~Ciafaloni and G.~Veneziano,
``Superstring Collisions At Planckian Energies,''
\hspi{1715674}{Phys.\ Lett.\ B {\bf 197} (1987) 81};
``Classical And Quantum Gravity Effects From Planckian Energy Superstring Collisions,''
\hspi{1778463}{Int.\ J.\ Mod.\ Phys.\ A {\bf 3} (1988) 1615}.



\bibitem{gross1}
D.~J.~Gross and P.~F.~Mende,
``The High-Energy Behavior Of String Scattering Amplitudes,''
\hspi{1707604}{Phys.\ Lett.\ B {\bf 197} (1987) 129};
``String Theory Beyond The Planck Scale,''
\hspi{1768484}{Nucl.\ Phys.\ B {\bf 303} (1988) 407};

\bibitem{gross2}
D.~J.~Gross,
``High-Energy Symmetries Of String Theory,''\\
\hspi{1801490}{Phys.\ Rev.\ Lett.\  {\bf 60} (1988) 1229.}




\bibitem{ooguri} P.~F.~Mende and H.~Ooguri,
``Borel Summation Of String Theory For Planck Scale Scattering,''
\hspi{2101629}{Nucl.\ Phys.\ B {\bf 339} (1990) 641}.






\bibitem{wittenn}
E.~Witten,
``Baryons In The 1/N Expansion,''
\hspi{348082}{Nucl.\ Phys.\ B {\bf 160} (1979) 57}.



\bibitem{hoof}
G.~'t Hooft,
``A Planar Diagram Theory For Strong Interactions,''\\
\hspi{19682}{Nucl.\ Phys.\ B {\bf 72} (1974) 461}.


\bibitem{holog1}
G.~'t Hooft,
``Dimensional Reduction In Quantum Gravity,''
\hre{gr-qc}{9310026}.




\bibitem{holog2}
L.~Susskind,
``The World as a hologram,''
J.\ Math.\ Phys.\  {\bf 36} (1995) 6377\\
\hre{hep-th}{9409089}.

\bibitem{sundrum1}
R.~Sundrum,
``Towards an effective particle-string resolution 
of the cosmological  constant problem,''
JHEP {\bf 9907} (1999) 001; \hre{hep-ph}{9708329}.


\bibitem{sundrum2}
``Fat gravitons, the cosmological constant and sub-millimeter
tests,''Phys.\ Rev.\ D {\bf 69} (2004) 044014;
\hre{hep-th}{0306106}.


\bibitem{zee}
A.~Zee,
``Dark energy and the nature of the graviton,''
\hre{hep-th}{0309032}.



\bibitem{deconstr}
N.~Arkani-Hamed, A.~G.~Cohen and H.~Georgi,
``(De)constructing dimensions,''
Phys.\ Rev.\ Lett.\  {\bf 86} (2001) 4757
\hre{hep-th}{0104005}.

\bibitem{c=11}
M.~R.~Douglas and S.~H.~Shenker,
``Strings In Less Than One-Dimension,''\\
\hspi{2078210}{Nucl.\ Phys.\ B {\bf 335} (1990) 635}.

\bibitem{c=12}
E.~Brezin and V.~A.~Kazakov,
``Exactly Solvable Field Theories Of Closed Strings,''
\hspi{2192420}{Phys.\ Lett.\ B {\bf 236} (1990) 144}.

\bibitem{c=13}
D.~J.~Gross and A.~A.~Migdal,
``Nonperturbative Two-Dimensional Quantum Gravity,''
\hspi{2078333}{Phys.\ Rev.\ Lett.\  {\bf 64} (1990) 127}.

\bibitem{matrix1}
T.~Banks, W.~Fischler, S.~H.~Shenker and L.~Susskind,
``M theory as a matrix model: A conjecture,''
Phys.\ Rev.\ D {\bf 55} (1997) 5112
\hre{hep-th}{9610043}.

\bibitem{matrix2}
N.~Ishibashi, H.~Kawai, Y.~Kitazawa and A.~Tsuchiya,
``A large-N reduced model as superstring,''
Nucl.\ Phys.\ B {\bf 498} (1997) 467
\hre{hep-th}{9612115}.






\bibitem{polyakov} A.~M.~Polyakov,
``The wall of the cave,''
Int.\ J.\ Mod.\ Phys.\ A {\bf 14} (1999) 645
\hre{hep-th}{9809057}.



\bibitem{polchinski} J.~Polchinski and M.~J.~Strassler,
``Hard scattering and gauge/string duality,''
Phys.\ Rev.\ Lett.\  {\bf 88} (2002) 031601
\hre{hep-th}{0109174}.




\bibitem{probe1}
T.~Banks, M.~R.~Douglas, G.~T.~Horowitz and E.~J.~Martinec,
``AdS dynamics from conformal field theory,''
\hre{hep-th}{9808016}.

\bibitem{probe2}
V.~Balasubramanian, P.~Kraus, A.~E.~Lawrence and S.~P.~Trivedi,
``Holographic probes of anti-de Sitter space-times,''
Phys.\ Rev.\ D {\bf 59} (1999) 104021\\
\hre{hep-th}{9808017}.

\bibitem{probe3}
P.~Kraus, F.~Larsen and S.~P.~Trivedi,
``The Coulomb branch of gauge theory from rotating branes,''
JHEP {\bf 9903} (1999) 003
\hre{hep-th}{9811120}.

\bibitem{probe4}
S.~R.~Das,
``Holograms of branes in the bulk and acceleration terms in SYM effective  action,''
JHEP {\bf 9906} (1999) 029
\hre{hep-th}{9905037}.


  \bibitem{non-ads1}
I.~R.~Klebanov and M.~J.~Strassler,
``Supergravity and a confining gauge theory: Duality cascades and  chiSB-resolution of naked singularities,''
JHEP {\bf 0008} (2000) 052
\hre{hep-th}{0007191}.

\bibitem{non-ads2}
J.~M.~Maldacena and C.~Nunez,
``Supergravity description of field theories on curved manifolds and a no  go theorem,''
Int.\ J.\ Mod.\ Phys.\ A {\bf 16} (2001) 822\\
\hre{hep-th}{0007018};
``Towards the large N limit of pure N = 1 super Yang Mills,''
Phys.\ Rev.\ Lett.\  {\bf 86} (2001) 588
\hre{hep-th}{0008001}.



\bibitem{duff}
M.~J.~Duff,
``Twenty years of the Weyl anomaly,''
Class.\ Quant.\ Grav.\  {\bf 11} (1994) 1387
\hre{hep-th}{9308075}.

\bibitem{bd} N. D. Birrell and P.C.W Davis, "{\em Quantum Fields
in Curved Sopacetime}", Cambridge Monographs on Mathemetical
Physics, CUP, 1982.


 \bibitem{Charmousis}
  C.~Charmousis and J.~F.~Dufaux,
  ``General Gauss-Bonnet brane cosmology,''
  Class.\ Quant.\ Grav.\  {\bf 19} (2002) 4671
  \hre{hep-th}{0202107}.
  
 


\end{thebibliography}
\end{document}